\documentclass[useAMS,usenatbib]{mn2e}

\usepackage{latexsym}
\usepackage{color}
\usepackage{amsmath}
\usepackage{amssymb}
\usepackage{graphicx}
\usepackage{aas_macros}
\newcommand{\Ms}{M$_{\odot}$}

\title[Streaming velocities at first star formation]{The Influence of
Streaming Velocities on the Formation of the First Stars}
\author[Schauer et al.]{Anna T. P. Schauer$^{1,2}$
\thanks{anna.schauer@utexas.edu} 
\thanks{NHFP Hubble Fellow},
Simon C. O. Glover$^{2}$, Ralf S. Klessen$^{2,3}$, Daniel Ceverino$^{2}$ \\
$^{1}$ Department of Astronomy, The University of Texas at Austin, Austin, TX 78712, USA\\
$^{2}$ Universit\"at Heidelberg, Zentrum f\"ur Astronomie, Institut f\"ur Theoretische
Astrophysik, Albert-Ueberle-Str. 2, 69120 Heidelberg, Germany\\
$^{3}$ Universit\"{a}t Heidelberg, Interdiszipli\"{a}res Zentrum f\"{u}r Wissenschaftliches Rechnen, Im Neuenheimer Feld 205, 69120 Heidelberg, Germany}

\begin{document}


\pagerange{\pageref{firstpage}--\pageref{lastpage}} \pubyear{2002}

\maketitle

\label{firstpage}

\begin{abstract}
How, when and where the first stars formed are fundamental questions regarding the epoch of Cosmic Dawn. A second order effect in the fluid equations was recently found to make a significant contribution: an offset velocity between gas and dark matter, the so-called streaming velocity.   Previous simulations of a limited number of low-mass dark matter haloes suggest that this streaming velocity can delay the formation of the first stars and decrease halo gas fractions and the halo mass function in the low mass regime. However, a systematic exploration of its effects in a large sample of haloes has been lacking until now. In this paper, we present results from a set of cosmological simulations
of regions of the Universe with different streaming velocities performed with the moving mesh code \textsc{arepo}. Our simulations have very high mass resolution, 
enabling us to accurately resolve minihaloes as small as $10^5 \: {\rm M_{\odot}}$. We show that in the absence of streaming, the least massive halo that contains cold gas has a mass $M_{\rm halo, min} = 5 \times 10^{5} \: {\rm M_{\odot}}$, but that cooling only becomes efficient in a majority of haloes for halo masses greater than $M_{\rm halo,50\%} = 1.6 \times 10^6 \: {\rm M_{\odot}}$. In regions with non-zero streaming velocities, $M_{\rm halo, min}$ and $M_{\rm halo,50\%}$ both increase significantly, by around a factor of a few for each one sigma increase in the value of the local streaming velocity. As a result, in regions with streaming velocities $v_\mathrm{stream} \ge 3\,\sigma_\mathrm{rms}$, cooling of gas in minihaloes is completely suppressed, implying that the first stars in these regions form within atomic cooling haloes. 
\end{abstract}

\section{Introduction}
%
The first stars in the Universe formed in dark matter minihaloes at high redshift, with masses of $M_\mathrm{minihalo} \approx 10^{5}--10^{7}$\,\Ms. Metal-free gas falling into these minihaloes was heated up by shocks and adiabatic compression. In the absence of cooling, the gravitational collapse of the gas would soon have stopped, with the gas becoming fully pressure supported. Previous studies have shown that in order to be able to avoid this fate, the gas must be able to form enough molecular hydrogen (H$_{2}$) to provide efficient cooling on a timescale short compared to the Hubble time \citep[see e.g.\ the reviews by][and references therein]{volkerreview13,glov13}.  Simple models suggest that the amount of H$_{2}$ formed in a minihalo is an increasing function of the minihalo mass, while the amount of H$_2$ required for efficient cooling decreases with minihalo mass \citep{tegmark97}. Consequently, there should be some minimum minihalo mass, $M_{\rm halo, min}$, marking the division between low-mass minihaloes in which H$_{2}$ cooling is inefficient and the gas does not form stars, and more massive minihaloes in which H$_{2}$ cooling is efficient and Population III (Pop~III) star formation can proceed. 

The value of this minimum halo mass is a crucial quantity for understanding the primordial Universe. In $\Lambda$CDM models, the comoving number density of dark matter haloes increases rapidly with decreasing halo mass $M_\mathrm{halo}$. Therefore, the number density of minihaloes capable of hosting Pop III star formation depends strongly on $M_{\rm halo, min}$, and consequently so does the Pop III star formation rate density at early epochs. This has several important implications.

Firstly, it implies that the size of the contribution that minihaloes make to cosmic reionisation may depend strongly on $M_{\rm halo, min}$ \citep[see e.g.][]{ahn12}. Pop III star-forming minihaloes are important sources of ionizing photons at high redshifts, while minihaloes that do not form Pop III stars play an important role by absorbing ionizing photons \citep{ham01,barkana02,cssf06}. 
Secondly, the metal pollution history of the Universe also depends on the minimum halo mass for Pop III star formation. Supernovae can efficiently eject large quantities of metal-enriched gas from low-mass minihaloes (\citealt{mml99,wise12,smith15}, see however \citealt{chiaki18}), and so a lower $M_{\rm halo, min}$ implies more widespread early metal enrichment of the intergalactic medium.
Finally, the value of $M_{\rm halo, min}$ also has important implications for possible observational tracers of the Pop~III epoch, such as the high redshift 21~cm background \citep{saai06,ycsc09} or the Pop~III supernova rate \citep{mattis16}.

However, despite the importance of $M_{\rm halo, min}$, there remains a surprising degree of uncertainty regarding its value. Efforts have been made to determine it using simplified toy models \citep[e.g.][]{tegmark97,ts09,glov13}, numerical simulations of individual minihaloes \citep[e.g.][]{fc00,met01} or numerical simulations of large populations of minihaloes \citep{yahs03}, but the scatter between the values obtained by different studies remains significant, as we examine in more detail in Section~\ref{sec:stream_ana1} below. 

In addition, most studies of $M_{\rm halo, min}$ have assumed that the baryons and the dark matter are initially at rest with respect to one another. However, it has recently been realized that in general this will not be the case, owing to the existence of residual velocity fluctuations in the baryonic component dating from the epoch when the baryons and photons were strongly coupled \citep{th10,tbh11}. These residual velocity fluctuations result in a systematic motion of the baryons relative to the dark matter prior to the onset of non-linear structure formation. They are 
coherent on a scale of several comoving Mpc (cMpc) and have a root-mean-squared value  of $\sigma_\mathrm{rms} \approx$~30~km~s$^{-1}{}$ at $z\approx1100$.

Although this relative velocity decreases as $v_\mathrm{stream} \sim (1+z)$ as the Universe expands, it nevertheless has profound effects on the formation of dark matter minihaloes. Streaming of baryons with respect to the dark matter leads to a reduced baryon fraction in the haloes \citep{naoz12} and a lower halo number density \citep{tbh11,naoz13}, as it is harder for baryons to settle into the host dark matter haloes. As a consequence, Pop~III star formation is delayed (\citealt{greif11,fbth12,oleary12,hirano18}, although see also \citealt{stacy11a} for a dissenting view). 

Simulations of the impact of streaming on the cooling and collapse of gas in individual minihaloes suggest that it leads to an increase in $M_{\rm halo, min}$. If true, this will have a clear impact on the high redshift 21~cm background \citep{fialkov14,fialkovreview14} and in extreme regions (i.e.\ regions with streaming velocities a few times greater than the rms value), it may also help to create the environment required for the formation of direct collapse black holes \citep{tm14,lns14,anna17b}. However, as yet no numerical studies have quantified the relationship between $v_{\rm stream}$ and $M_{\rm halo, min}$ for a large sample of minihaloes over a broad range of redshifts. It is this lack that we attempt to remedy in our current paper.

We present here the results of a set of high resolution cosmological simulations carried out with streaming velocities $v_{\rm stream} = 0, 1, 2$, and 3 times $\sigma_\mathrm{rms}$. Several thousand minihaloes form in each simulation by our final redshift $z=14$, allowing us to examine the impact of the streaming on a large statistical sample of minihaloes.
In contrast to previous studies \citep[e.g. ][]{naoz12,naoz13,pnmv16}, we also include a primordial chemistry network in our simulations, allowing us to investigate the thermal evolution of the gas. We can therefore not only investigate the gas fraction in our studies, but also quantify how much cold, dense gas is available for Pop~III star formation in the haloes. In most of our simulations, we use a very high resolution of 20\,\Ms\; per gas cell, allowing us to follow the collapse of the gas up to densities of $n > 10^2$\,cm$^{-3}$ in every minihalo in which cooling and collapse takes place. 
We are therefore able to derive quantitative and statistically robust results
about the influence of streaming velocities on first star formation. 

Our paper is structured as follows: we present our set of simulations in 
Section~\ref{sec:stream_meth}. In Section~\ref{sec:stream_ana}, we analyse the results of our study, starting with the simulation without streaming velocities in Section~\ref{sec:stream_ana1}. In Section~\ref{sec:stream_ana2}, we then present the results for simulations with streaming velocities of 1, 2 or 3\,$\sigma_\mathrm{rms}$. 
We give our conclusions in Section~\ref{sec:stream_con}. 
%
\section{Method}
\label{sec:stream_meth}
\subsection{Numerical method}
Our cosmological simulations are carried out using the moving-mesh code {\sc arepo}
\citep{arepo}. {\sc arepo} solves the equations of hydrodynamics on an unstructured mesh
defined by the Voronoi tessellation of a set of mesh-generating points that move with the flow
of the gas. Dark matter is included using a \citet{bh86} oct-tree. 

To model the chemical  and thermal evolution of the gas, we use the same primordial chemistry network and cooling function as in \citet{anna17b}. These are based on the versions implemented in {\sc arepo} by \citet{hartwig15a}, but we have updated them in several respects. Most significantly, we have included a simplified model of deuterium chemistry, focused on the formation and destruction of hydrogen deuteride, HD. In addition, we have updated several of the chemical rate coefficients used within the model. Full details regarding these updates are given in Appendix~\ref{app:chem}.

\subsection{Initial conditions}
Our simulations are initialised at $z=200$.  
Their details are summarised in Table \ref{tab:stream_sim}.
The initial conditions for the dark matter are created with MUSIC \citep{hahn11},  
using the transfer functions of \cite{eh98}. 
The baryons are assumed to initially trace the dark matter density distribution. 
We assume a $\Lambda$CDM cosmology and use cosmological parameters derived from {\sc planck} 
observations of the CMB, namely $h = 0.6774$,  
$\Omega_0 = 0.3089$, $\Omega_\mathrm{b} = 0.04864$, $\Omega_\Lambda = 0.6911$,  
$n = 0.96$ and $\sigma_8 = 0.8159$ \citep{Planck15}.

When setting up our initial conditions, we neglect differences in the density fluctuations 
for gas and dark matter. This can lead to increased baryon fractions in minihaloes 
and to a lower the minimum halo mass
\citep[see e.g.][]{naoz05,naoz09,naoz11}. However, for our models the effect is less than $<$1\% 
compared to work by \citealt{pnmv16} who treat this effect more carefully. 
To do better, a more sophisticated treatment is necessary, e.g. 
by using codes such as  BCCOMICS \citep{ahn18} 
or creating transfer functions based on a modified version of CMBFAST \citep{cmbfast96,pnmv16} 
for setting up initial conditions.

In our runs without streaming, the initial velocity field of the baryons is the same as that
of the dark matter. In the runs that include streaming, we add a constant velocity offset, 
arbitrarily chosen to be in the $x$-direction, with a magnitude $v_{\rm stream}$. 
We run simulations with $v_{\rm stream} = 6, 12,$ and 18~km~s$^{-1}$, corresponding 
to values 1, 2 and 3 times $\sigma_{\rm rms}$, the root-mean-squared streaming
velocity at $z = 200$.
In our initial conditions, we ignore the effect of a position 
shift of the baryons with respect to the dark matter due to the streaming, as \citet{naoz12}
have shown that this is unimportant at $z \gg 15$ and is completely erased by the time $z \sim 15$.

The naming convention for our simulations is simple. 
Each simulation has a name of the form v$N$, where $N$ denotes the streaming velocity 
in units of $\sigma_{\rm rms}$; thus, v0 corresponds to a run with $v_{\rm stream} = 0$.
Most of the simulations are carried out with a box size of (1\,cMpc/$h$)$^3$. 
However, for the case with $v_{\rm stream} = 3 \, \sigma_{\rm rms}$, 
we have carried out an additional simulation with a significantly larger box size, $($4Mpc/$h)^3$, which we name v3\_big.  

All simulations are performed with 1024$^3$ dark matter particles and 
initially have 1024$^3$ Voronoi mesh cells to represent the gas. This corresponds 
to a dark matter particle mass of $99 \, {\rm M_{\odot}}$ and an average initial mesh 
cell mass of $18.6 \, {\rm M_{\odot}}$ in the simulations with a box size of $($1Mpc/$h)^3$. We have verified that this mass resolution is sufficient to obtain numerically converged results, as discussed in more detail in Appendix~\ref{converge}. The corresponding values in the run with a larger box are larger by a factor of $4^{3} = 64$.

We use ths standard {\sc arepo} refinement scheme, in which the code refines or de-refines cells as necessary to keep the mass of the gas close to the initial cell mass. At high densities, we prevent run-away collapse by switching off refinement once the cell volume drops below $0.1 \: h^{-3} \, {\rm cpc^{3}}$. The corresponding maximum density is 
redshift dependent, but is typically a few times $10^{6} \: {\rm cm^{-3}}$. Since we are interested in this paper in the behaviour of gas at densities orders of magnitude lower than this value, this choice does not affect our results.
\begin{table}
\begin{center}
\begin{tabular}{l|cccc}
Name  & Box size       & $v_\mathrm{stream}$ & $M_\mathrm{gas}$  & $M_\mathrm{DM}$ \\
      & [cMpc$/h$]     & [km\,s$^{-1}$]      & [\Ms]               & [\Ms]             \\
\hline
v0     & 1  &  0 &  18.6 & 99  \\
v1     & 1  &  6 &  18.6 & 99  \\
v2     & 1  & 12 &  18.6 & 99  \\
v3     & 1  & 18 &  18.6 & 99  \\
v3\_big& 4  & 18 &  1190 & 6360  \\
\hline
\end{tabular}
\caption[Overview of our simulations.]
{Overview of our simulations. Note that $v_{\rm stream}$ gives 
the strength of the streaming velocity at $z=200$. $M_{\rm gas}$ and $M_{\rm DM}$ are the initial mesh cell gas mass and dark matter particle mass, respectively.
\label{tab:stream_sim}}
\end{center} 
\end{table}
%
\subsection{Halo selection}
We select haloes via a standard friends-of-friends algorithm.
We have verified that our results do not change significantly if we look only at the first subhalo in each halo that is actually gravitationally bound (see Appendix~\ref{app:fof}).

At each output redshift, we examine every halo and identify those containing cold gas.
In order to be counted as ``cold and dense'' for our purposes, gas needs to fulfill the following
three criteria:
\begin{itemize}
\item Temperature $T \leq 500$~K,
\item Number density $n\geq 100$~cm$^{-3}$,
\item Fractional H$_{2}$ abundance  $x_{\rm H_{2}} \geq 10^{-4}$.
\end{itemize}
These criteria allow us to distinguish between gas which has formed H$_{2}$,
cooled and begun to undergo gravitational collapse, and gas which instead has
a low temperature simply because it has not yet virialized. Our selection is
insensitive to the values chosen for the temperature and molecular hydrogen
abundance criteria, provided we choose a temperature lower than the typical
halo virial temperature and larger than the minimum temperature reachable by
H$_{2}$ cooling, $T \sim 200$~K, and an H$_{2}$ abundance that is considerably
larger than the abundance in the undisturbed intergalactic medium, $x_{\rm H_{2}}
\sim 10^{-6}$ \citep{hp06}. 
We assume that gas denser than 
our threshold has reached the ``point of no return'' beyond which the gas cannot be stopped 
from collapsing, implying that the halo will form stars shortly afterwards.
Any chosen density threshold is to some extent arbitrary. 
In fact, the minimum halo mass, which we investigate in Section \ref{sec:stream_ana}, does change by a few tens of percent for lower or higher 
number densities. 
We define the cold mass, $M_\mathrm{cold}$, as the sum of the gas masses in all of the
mesh cells that fulfill the three criteria above.  
A halo that contains at least one such gas cell is counted to be ``cold''
and contributes to the number of cold gas haloes, $N_\mathrm{cold}$.
\section{Results}
\label{sec:stream_ana}
\subsection{No streaming velocities}
\label{sec:stream_ana1}
\begin{figure}
\centering
\includegraphics[width=0.99\columnwidth]{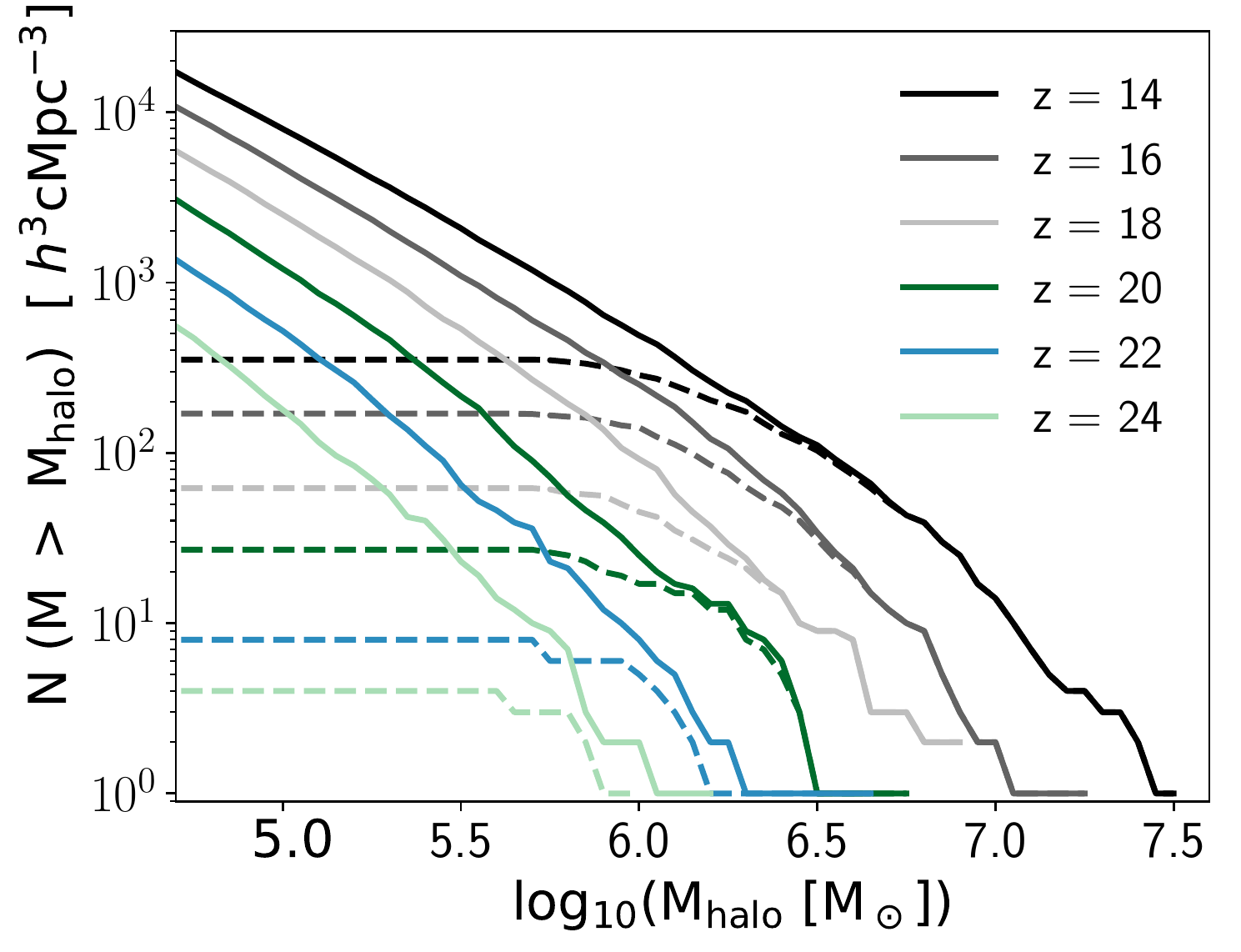}
\caption[Cumulative halo mass function for all haloes and cold haloes 
for the simulation without streaming velocity.]
{Cumulative halo mass function for all haloes (solid lines) and 
cold haloes (dashed lines) for redshifts from $z=14$ to $z=24$ 
for the simulation without streaming. }
\label{fig:pss-z}
\end{figure}
We begin by analysing the evolution of haloes 
in a patch of the Universe with zero streaming velocity.  
Figure~\ref{fig:pss-z} shows the cumulative halo mass function $N (> M)$ 
plotted for several different redshifts in the range $14 \le z \le 24$. 
We also show the cumulative cold halo mass function $N_\mathrm{cold} (> M)$ for all 
haloes that contain at least one cold, dense gas cell.
At redshifts $z \le 20$, the most massive haloes are always cold, whereas smaller haloes with masses less than a few times $10^{5}$\;\Ms\, are never cold, independent of redshift. 
\begin{figure}
\centering
\includegraphics[width=0.99\columnwidth]{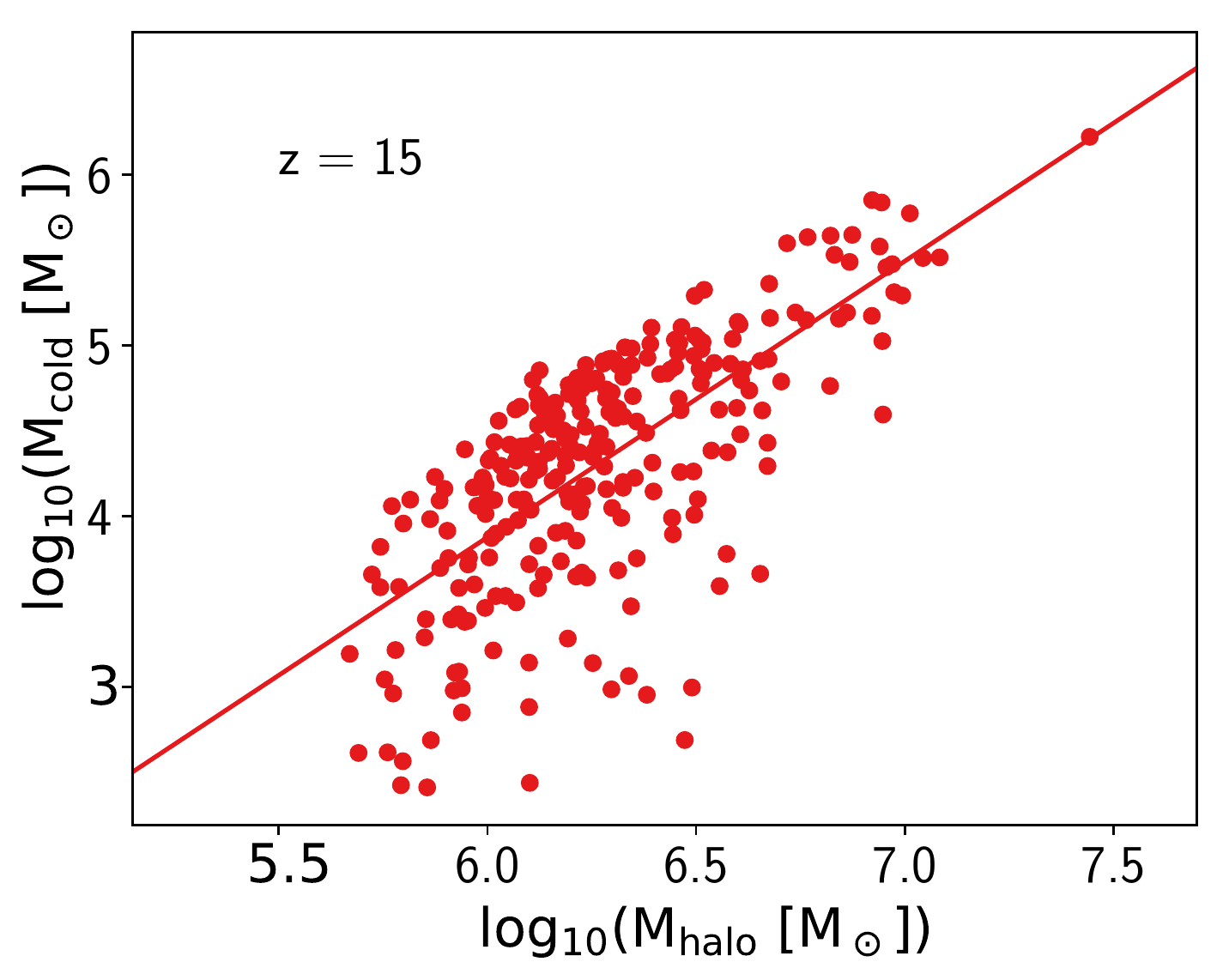}
\caption[Cold gas mass $M_\mathrm{cold}$ plotted as a function of halo mass $M_\mathrm{halo}$ 
for the simulation without streaming velocity at $z=15$.]
{Cold gas mass $M_\mathrm{cold}$ plotted as a function of 
halo mass $M_\mathrm{halo}$ for the simulation without streaming velocity 
at $z=15$. The solid red line provides a fit to the data points.}
\label{fig:mhalo_mcold_z15}
\end{figure}

Figure \ref{fig:mhalo_mcold_z15} shows the cold gas mass $M_\mathrm{cold}$ 
plotted against the halo mass $M_\mathrm{halo}$ for 
all haloes that contain at least one cold gas cell at redshift $z = 15$. 
In most of the haloes, there is a clear correlation between cold gas mass and halo mass that can be fit with the function
\begin{equation}
M_\mathrm{cold} =  7.55 \times 10^3 \:\mathrm{M}_\odot \times \left( \frac{M_\mathrm{halo}}{10^6 \: \mathrm{M}_\odot}\right)^{1.6},
\end{equation}
indicated by the red line in the plot. The mass in stars is 
often inferred directly from the halo mass in semi-analytical models of Pop.\ III star formation \citep[see e.g.][]{hartwig15b,magg18}, and we 
provide the amount of cold gas available as a function of halo mass as an in between step. 
We have computed similar fits for all of the simulations that we have carried out (see Appendix~\ref{sec:lw2_relation}). However, it is clear from the Figure that this relationship is at best a crude approximation. In particular, there are a number of haloes in the mass range $10^{5.7} < M < 10^{6.5} \: {\rm M_{\odot}}$ that sit far below the fit, with cold gas masses $M_{\rm cold} < 10^{3} \: {\rm M_{\odot}}$. 

\begin{figure}
\centering
\includegraphics[width=0.99\columnwidth]{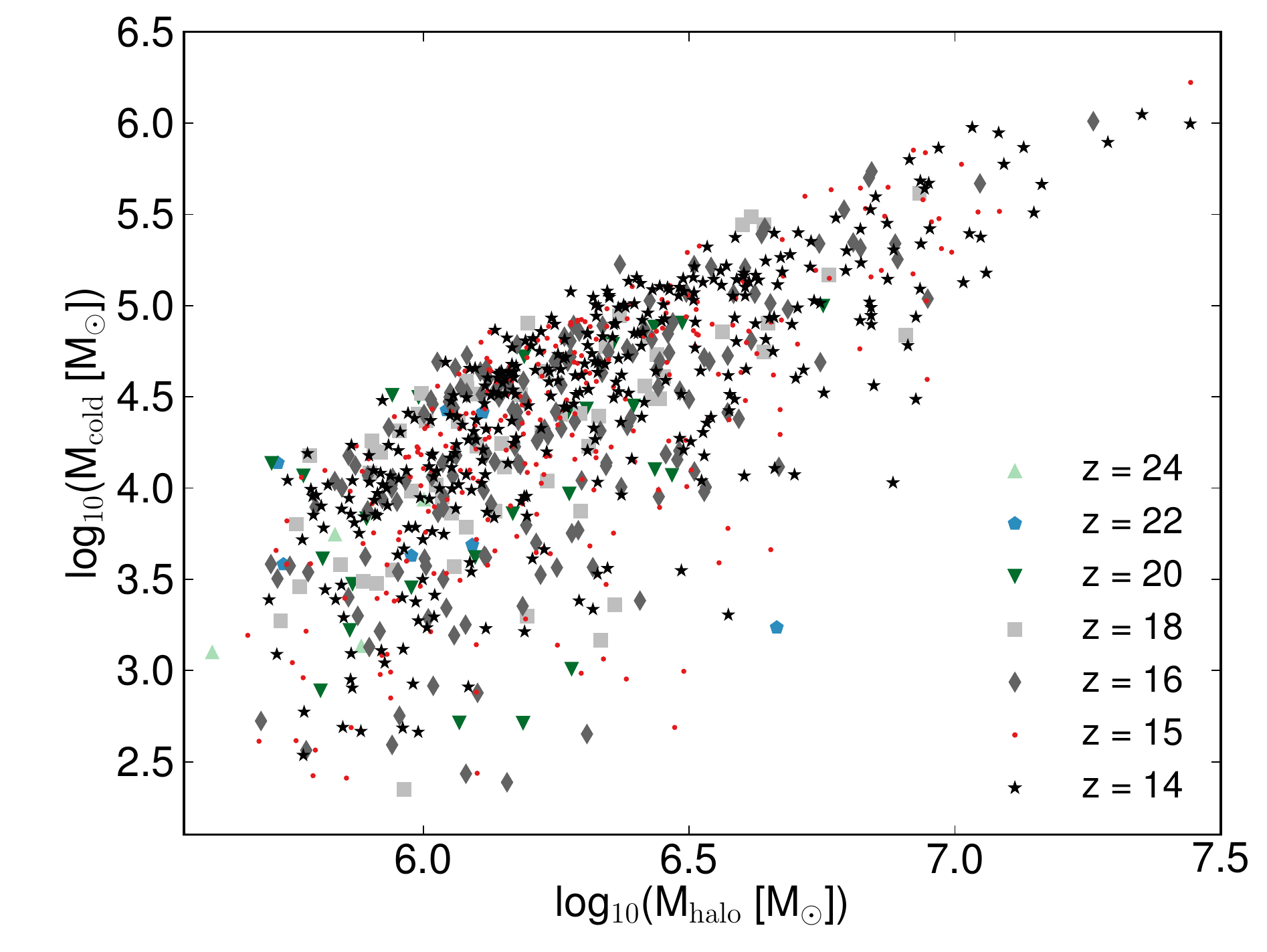}
\caption[Cold gas mass $M_\mathrm{cold}$ plotted as a function of halo mass 
$M_\mathrm{halo}$ for the simulation without streaming velocity at redshifts in the range $14 \le z \le 24$.]
{Cold gas mass $M_\mathrm{cold}$ plotted as a function of 
halo mass $M_\mathrm{halo}$ for the simulation without streaming velocity
at redshifts in the range $14 \le z \le 24$.}
\label{fig:mhalo_mcold_allz}
\end{figure}
The relation between the cold gas mass and the halo mass does not vary strongly with redshift over the range of redshifts studied here. Figure \ref{fig:mhalo_mcold_allz} shows the cold gas mass--halo mass relation  
for the same simulation for a number of different redshifts in the range $14 \le z \le 24$. At redshifts $z=24$ (light green triangles) and $z=22$ (blue pentagons), we are limited by small number statistics as our simulation includes only 4 and 8 haloes that contain cold gas, respectively. For redshifts $z \le 20$, the cold mass--halo mass relations for different redshifts agree well.

\begin{figure}
\centering
\includegraphics[width=0.99\columnwidth]{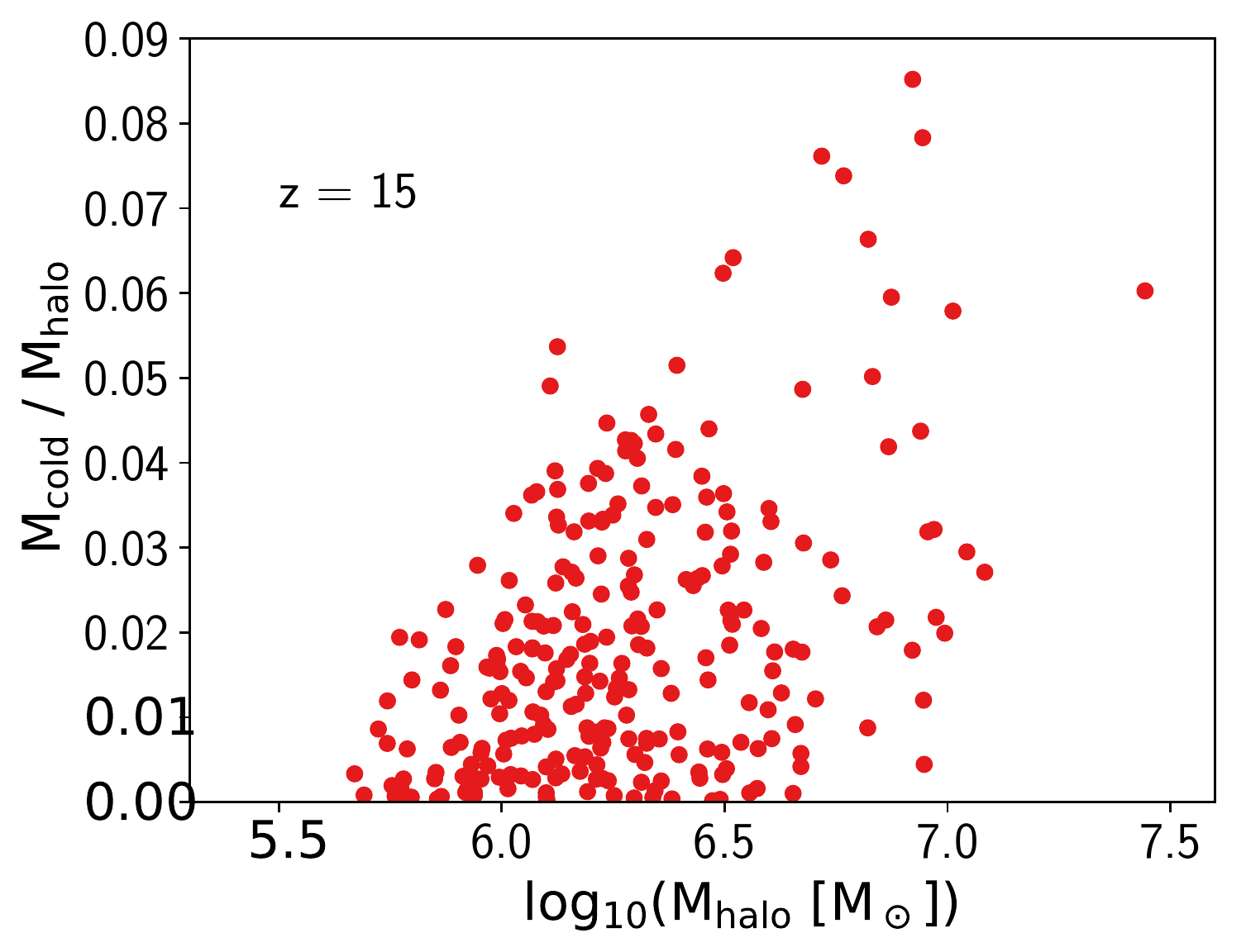}
\caption[Cold gas fraction $M_\mathrm{cold}/M_\mathrm{halo}$ plotted as a function
of  halo mass $M_\mathrm{halo}$ for the  simulation without streaming velocity at $z=15$.]
{Cold gas fraction $M_\mathrm{cold}/M_\mathrm{halo}$ plotted as a function 
of  halo mass $M_\mathrm{halo}$ for the 
simulation without streaming velocity at $z=15$. 
Values are plotted only for haloes that contain at least one 
cold gas cell. }
\label{fig:coldgasfraction}
\end{figure}

As expected, larger haloes contain more cold gas, but in part this is because they contain more gas in total. We can more easily examine whether cooling is more efficient in these haloes by plotting the cold gas 
fraction $M_\mathrm{cold}/M_\mathrm{halo}$\:as a function 
of halo mass $M_\mathrm{halo}$, as shown in Figure \ref{fig:coldgasfraction}. 
We see a weak correlation between cold gas fraction and halo mass. 
Larger haloes tend to have higher cold gas fractions, 
but the scatter is large. 

\begin{figure}
\centering
\includegraphics[width=0.99\columnwidth]{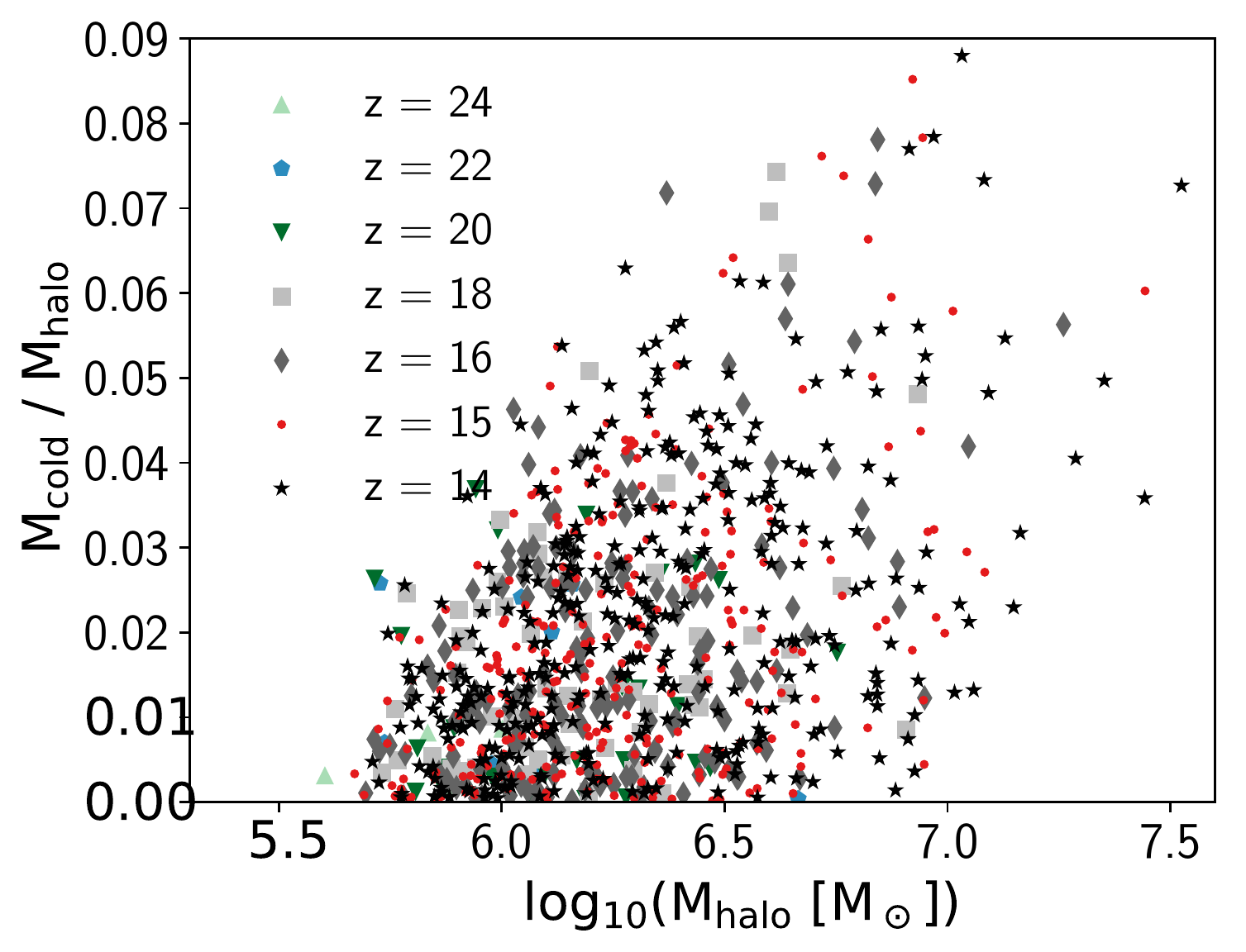}
\caption[Cold gas fraction $M_\mathrm{cold}/M_\mathrm{halo}$ plotted as a function
of  halo mass $M_\mathrm{halo}$ for the simulation without 
streaming velocity for redshifts in the range $14 \le z \le 24$.]
{Cold gas fraction $M_\mathrm{cold}/M_\mathrm{halo}$ plotted as a function  of  halo mass $M_\mathrm{halo}$ for the
simulation without streaming velocity
for redshifts in the range $14 \le z \le 24$. 
Values are plotted only for haloes that contain at least one             
cold gas cell.}
\label{fig:coldgasfraction_allz}
\end{figure}

The cold gas fraction is similar for higher redshifts as can be seen in 
Figure \ref{fig:coldgasfraction_allz}. The main difference arises from 
the fact that the haloes forming at higher redshifts are less massive 
and thus the diagram is less populated at higher halo masses for higher redshifts. 

\begin{figure}
\centering
\includegraphics[width=0.99\columnwidth]{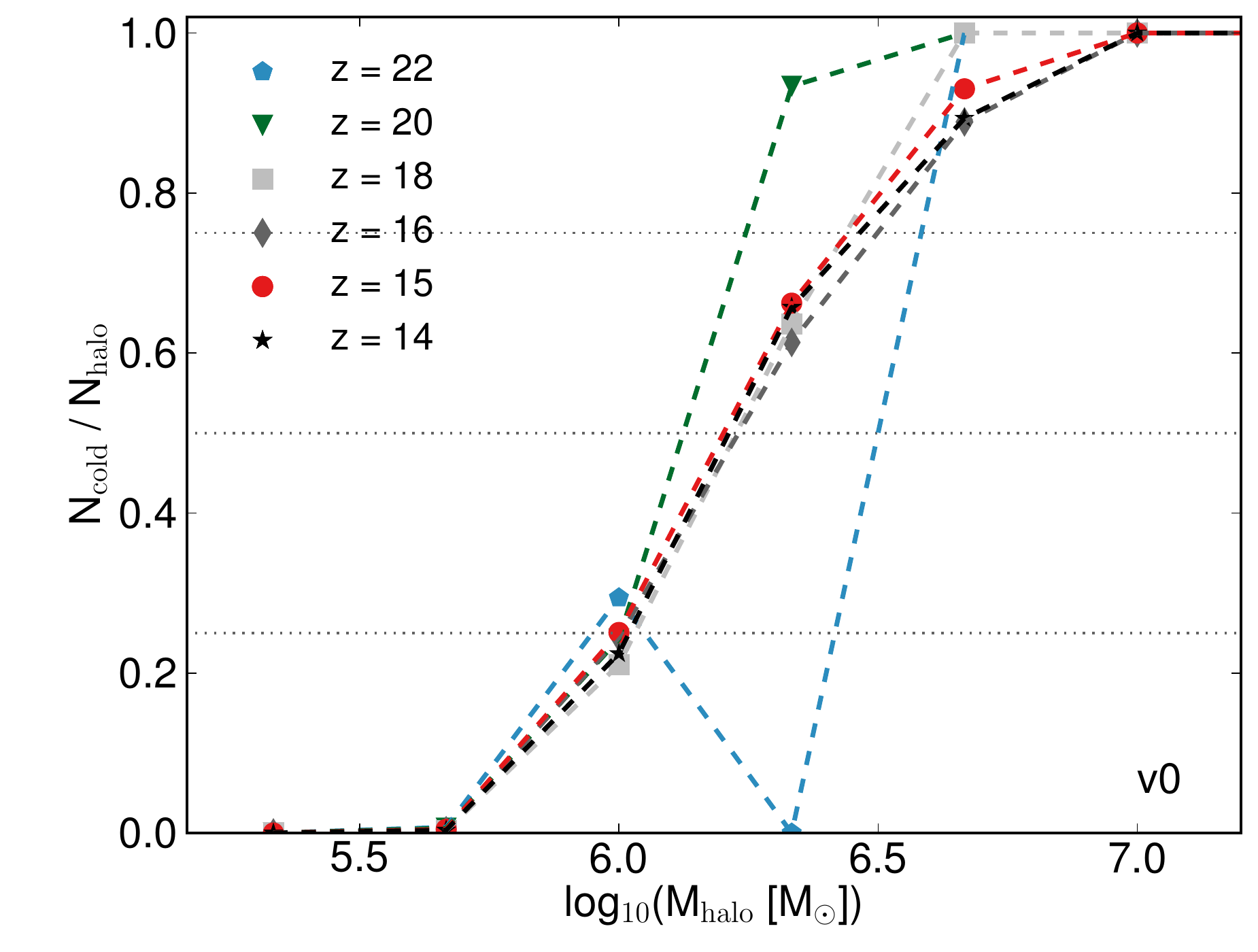}
\caption[Fraction of haloes that contain cold gas plotted as function of halo mass
for different redshifts in the range $14 \le z \le 22$.]
{Fraction of haloes of a given mass that contain cold gas plotted as function of halo mass for different redshifts in the range $14 \le z \le 22$. The dotted, grey horizontal lines mark the 25\%, 50\% and 75\% transitions of $N_\mathrm{cold} / N_\mathrm{halo}$ from fewer 
to more cold haloes. $N_\mathrm{cold}/N_\mathrm{all}$ generally 
increases with halo mass. At redshift $z=22$ we  
are limited by low number statistics, since only 8 haloes in the simulation contain any cold gas at this redshift.}
\label{fig:fraction}
\end{figure}

Another interesting quantity is the minimum halo mass $M_\mathrm{halo,min}$ required for haloes to contain cold gas. Since gas cooling and the consequent loss of thermal pressure support is a necessary condition for star formation in these low-mass haloes, this is also the minimum halo mass required for Pop.\ III star formation.
There appears to be a clear edge to the distribution of points on the left-hand side 
of Figures \ref{fig:mhalo_mcold_z15}, \ref{fig:mhalo_mcold_allz},
\ref{fig:coldgasfraction} and \ref{fig:coldgasfraction_allz} 
that corresponds to a minimum halo mass. 
To quantify this, 
we show the fraction of haloes with at least one cold gas 
particle, $N_\mathrm{cold}$, over the number of haloes in the mass bin, 
$N_\mathrm{all}$, for different redshifts in Figure \ref{fig:fraction}.
At all times, the fraction $N_\mathrm{cold}/N_\mathrm{all}$ increases with 
halo mass $M_\mathrm{halo}$. 
At redshifts $z \le 22$, there is always a minimum halo mass 
above which more than 50\% of all haloes harbour cold gas. The transition 
happens at $M_\mathrm{halo} \approx 1.6 \times 10^6$\,\Ms\;and 
is independent of redshift. 

\begin{figure}
\centering
\includegraphics[width=0.99\columnwidth]{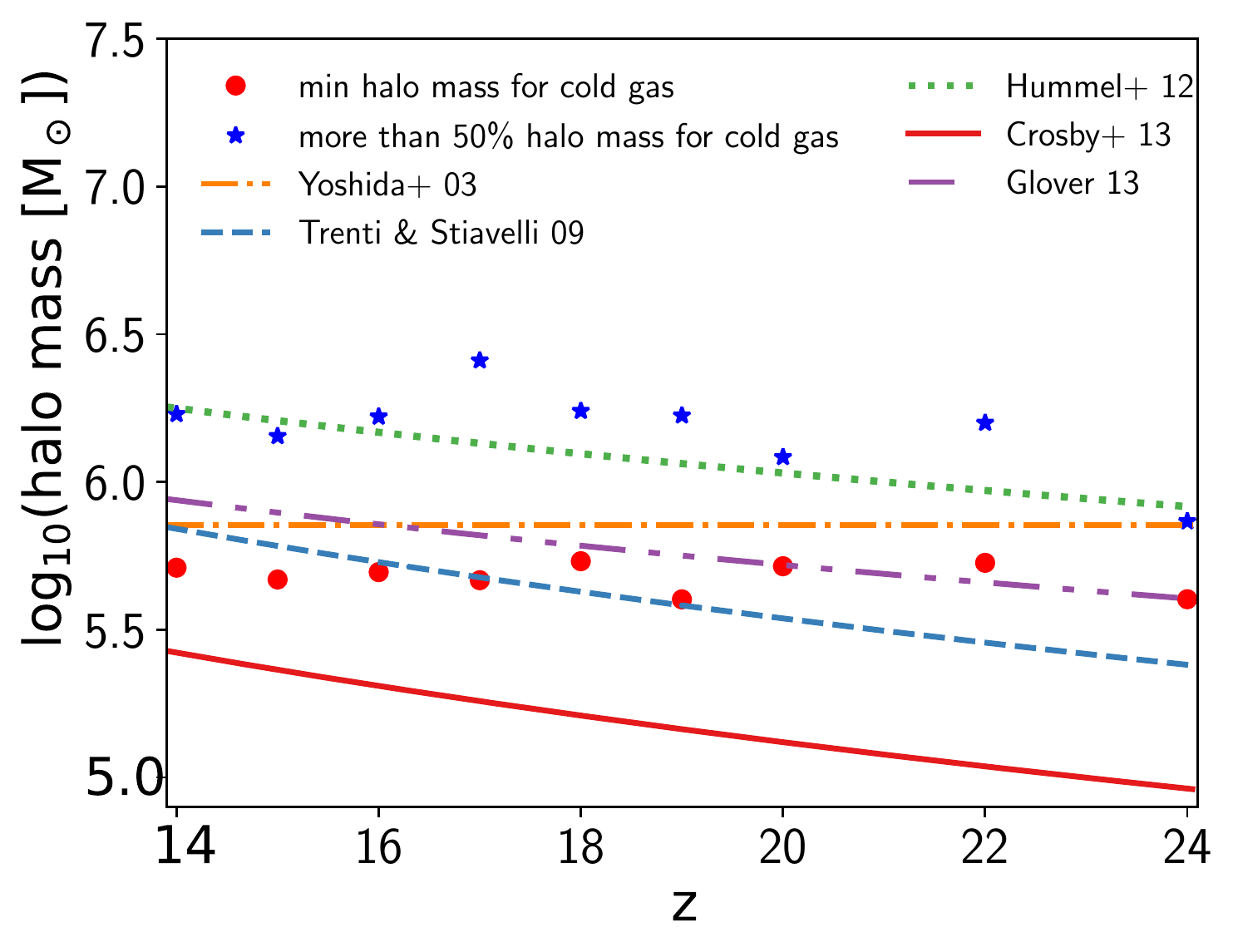}
\caption[Minimum halo mass $M_\mathrm{halo,min}$ for a halo to contain cold gas
and halo mass $M_\mathrm{halo,50\%}$ above which more than 50\% of all 
haloes harbour cold, dense gas plotted as a function of redshift. 
Data from the literature are included.]
{Minimum halo mass $M_\mathrm{halo,min}$ for a halo to contain cold gas 
and halo mass $M_\mathrm{halo,50\%}$ above which more than 50\% of all 
haloes harbour cold, dense gas plotted as a function of redshift. Over-plotted are data 
from the literature from \protect\cite{yahs03}, \protect\cite{ts09},
\protect\cite{hum12}, \protect\cite{crosby13}, and \protect\cite{glov13}. 
Note that the results from \protect\cite{yahs03} are empirical results from a cosmological simulation, while the other values are from simplified models of cooling and collapse in high redshift minihaloes.}
\label{fig:start}
\end{figure}
We show these two masses, the minimum halo mass $M_\mathrm{halo,min}$\; and the halo mass above which 50\% or more of haloes contain cold gas
$M_\mathrm{halo,50\%}$\; in Figure \ref{fig:start}. 
In addition, we also show several models from the literature for comparison. 

Simulations by \citet[orange dot-dashed line]{yahs03} 
find a minimum halo mass of $\approx 7 \times 10^5$\,\Ms, 
which is about 50\% larger than our value. 
As their simulation was carried out at lower resolution (their best resolution is about a factor of two 
lower than ours, $m_\mathrm{gas} \approx 42$\,\Ms), used an earlier and less accurate treatment of the H$_2$ cooling function, and also adopted a higher density threshold for their cold gas ($n \ge 5\times 10^2$\,cm$^{-3}$), this small discrepancy is not surprising. 

Other authors have used semi-analytic models to predict the 
minimum halo mass. Some of these models produce values very close to ours at the middle of our range of redshifts \citep[e.g.][]{ts09,glov13}, but it is clear that all of the semi-analytical models predict an inverse relationship between $M_\mathrm{halo,min}$\; and redshift that is not seen in our simulation or in the \citet{yahs03} results.

Finally, it is important to note that $M_\mathrm{halo,min}$
appears to be quite sensitive to density threshold used to identify haloes with cold, dense gas, changing by tens of percent for changes in the threshold of a factor of a few. $M_\mathrm{halo,50\%}$, on the other hand, has little sensitivity to the choice of threshold.

\subsection{Streaming velocities}
\label{sec:stream_ana2}
We now examine how increasing the streaming velocity affects the ability of gas to cool and form stars within low-mass haloes.
We start by looking at two effects that have previously been studied by other authors and that we can confirm with our own simulations. 

\begin{figure}
\centering
\includegraphics[width=0.99\columnwidth]{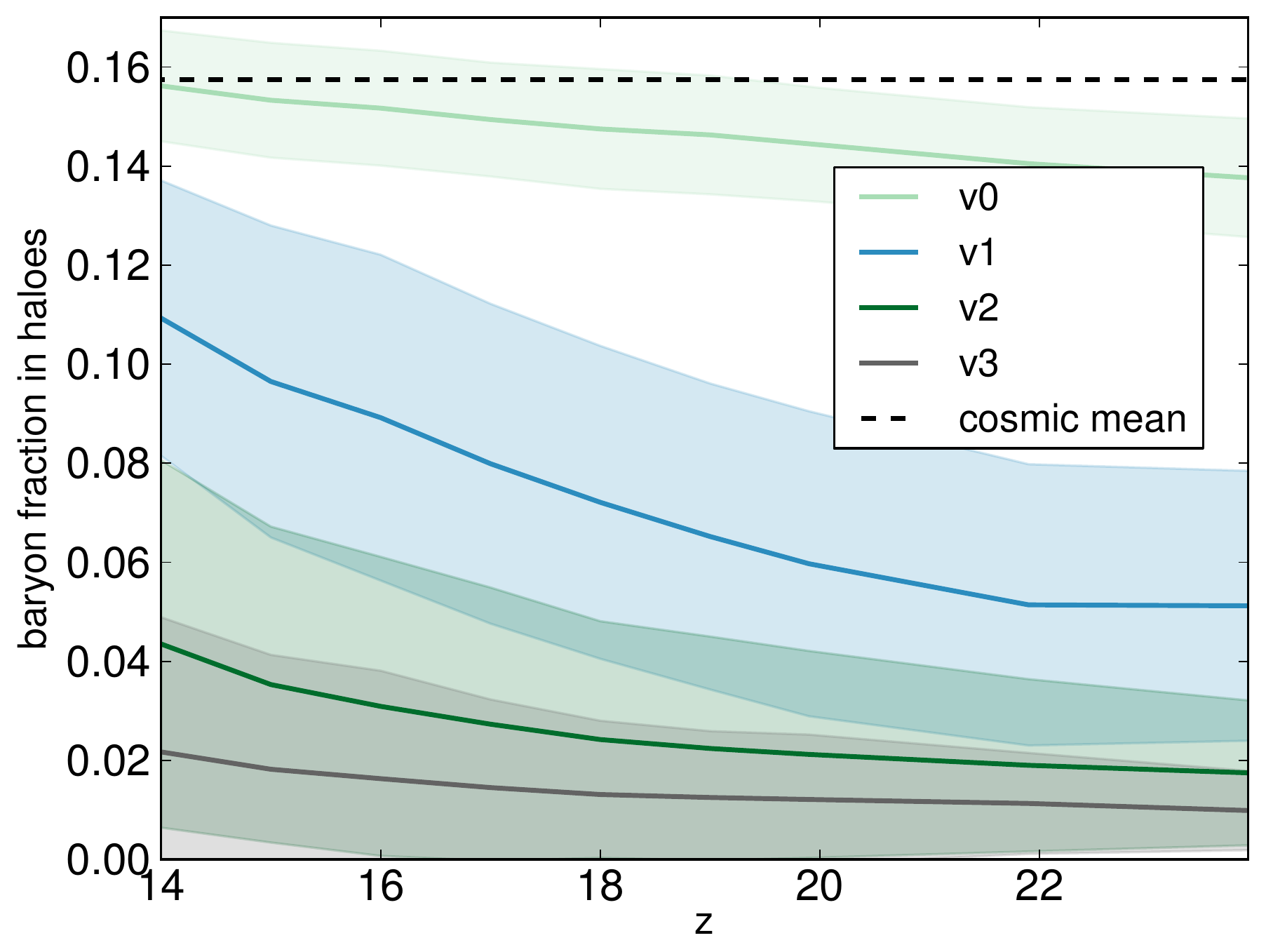}
\caption[Baryon fraction a function of redshift for different streaming velocity values.]
{Baryon fraction in haloes with masses 
of $M_\mathrm{halo} \ge 10^5$\,\Ms\;as a function of redshift. 
In the case of zero streaming velocity (light green), 
the value approaches the cosmic mean (dashed line). For all simulations 
with non-zero streaming velocities (blue, dark green, grey), it is well below that value. 
The shaded regions show the respective standard deviations.}
\label{fig:fbar}
\end{figure}

First, a non-zero streaming velocity makes it harder for baryons to settle into low-mass dark matter haloes. This can be quantified by examining the behaviour of the baryon fraction $M_\mathrm{bar}/ M_\mathrm{halo}$ as a function of redshift, as we do in Figure~\ref{fig:fbar}. Here we show the evolution with redshift of the mean and the standard deviation of the baryon fraction in haloes with mass $M > 10^5 \: {\rm M_{\odot}}$ or higher in runs v0, v1, v2 and v3. 
For the range of redshifts examined here, the Jeans mass $M_{\rm J} \ll 10^{5} \: {\rm M_{\odot}}$, and so in the absence of streaming we expect the baryon fraction to be close to the cosmic mean of $\Omega_\mathrm{b}/\Omega_0  \approx 0.16 $, as is the case in run v0. However, as we increase the streaming velocity, the baryon fraction decreases, reaching values as low as 0.01 at early times in the v3 run. 
This decrease in gas fraction was predicted by \cite{th10} and has been noted in previous simulations by several authors \citep[e.g.][]{tbh11,naoz13}. We confirm the effect with our own high resolution data. 

\begin{figure}
\centering
\includegraphics[width=0.99\columnwidth]{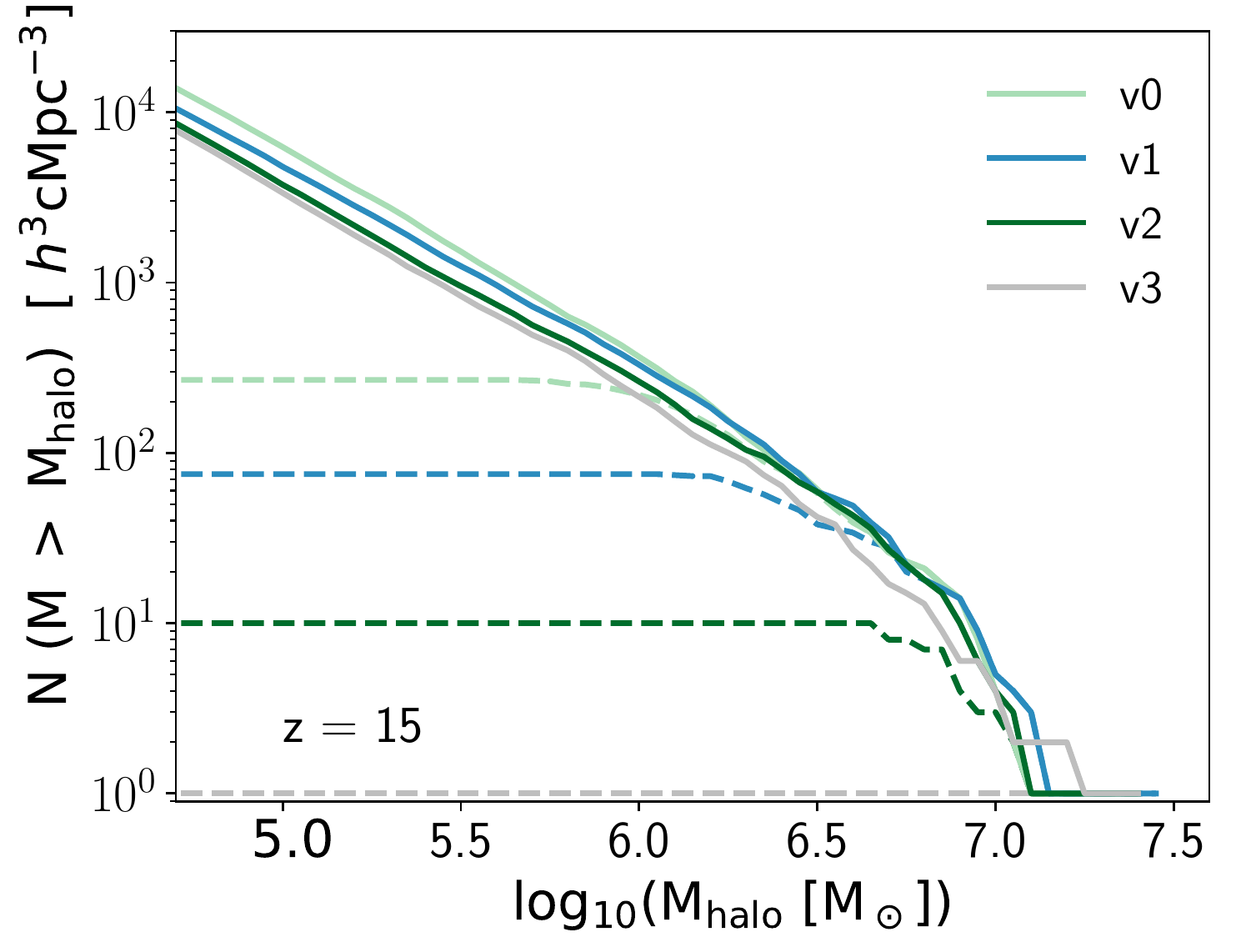}
\caption[Cumulative halo mass function $N(>M_\mathrm{halo})$ 
at redshift $z=15$ for all simulation in the small box and  
streaming velocities of 0, 1, 2 and 3 $\sigma_\mathrm{rms}$.]
{Cumulative halo mass function $N(>M_\mathrm{halo})$ 
at redshift $z=15$ for simulations v0, v1, v2 and v3. 
The total number of haloes is shown as a 
solid line and the number of haloes with cold gas as a dashed line.}
\label{fig:pss_vs}
\end{figure}

A second large-scale effect follows as a consequence: 
the halo mass 
function decreases in simulations of non-zero streaming velocity regions 
owing to the decreased contribution of baryons to the self-gravity 
of overdense regions. 
We show how the influence of streaming velocities lowers the 
cumulative halo mass function $N(>M_\mathrm{halo})$ at redshift $z=15$  in Figure \ref{fig:pss_vs} 
for all of the simulations carried out using the $($1Mpc/$h)^3$ simulation volume. The suppression in our simulations is on the order of 10\% to 25\% in the mass range 
of $M_\mathrm{halo} = 10^5 - 10^6$\,\Ms\;at $z=15$ for each $\sigma_\mathrm{rms}$ increase in the streaming 
velocity. 
\cite{naoz12} see a decrease by 15\% and 50\% for 1.7$\sigma_\mathrm{rms}$ 
and 3.4$\sigma_\mathrm{rms}$ regions, respectively at redshift $z=19$. 
\cite{pnmv16} find a similar decrease of small objects with 
$M_\mathrm{halo} \le 10^5$\,\Ms\;of $\sim$\:60\% at redshift $z=20$. 
Our results are therefore in good agreement with other recent studies.  

We also note that the overall normalisation of our halo mass function is larger than that in the run of \citet{naoz12} with a normal $\sigma_8 = 0.82$. This may be due to the different way in which we define halo masses compared to their study: they consider only gas and dark matter within the virial radius, while we consider the mass of the whole structure found by the friends-of-friend algorithm.  

\begin{figure}
\centering
\includegraphics[width=0.99\columnwidth]{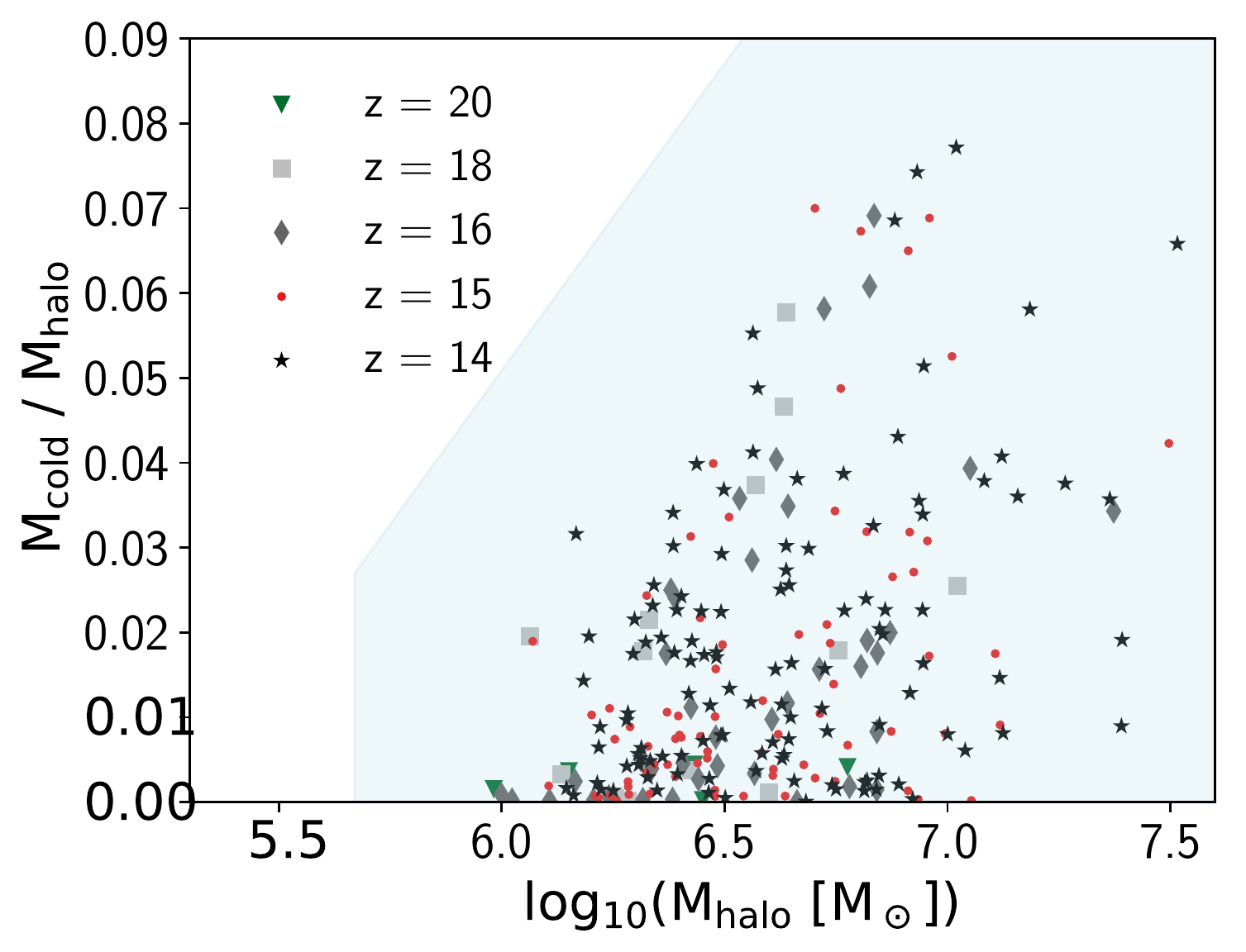}
\caption[Cold gas fraction $M_\mathrm{cold}/M_\mathrm{halo}$ as a function of 
halo mass for the simulation with a streaming velocity of 1$\sigma$ for 
redshifts in the range $14 \le z \le 20$.]
{Cold gas fraction $M_\mathrm{cold}/M_\mathrm{halo}$ as a function of 
halo mass for the simulation with a streaming velocity of 1$\sigma$ for 
redshifts in the range $14 \le z \le 20$. 
Values are only plotted for haloes that contain at least one cold gas cell. 
The shaded region shows the parameter space occupied by simulation v0.}
\label{fig:coldgasfraction_v06_allz}
\end{figure}

\begin{figure}
\centering
\includegraphics[width=0.99\columnwidth]{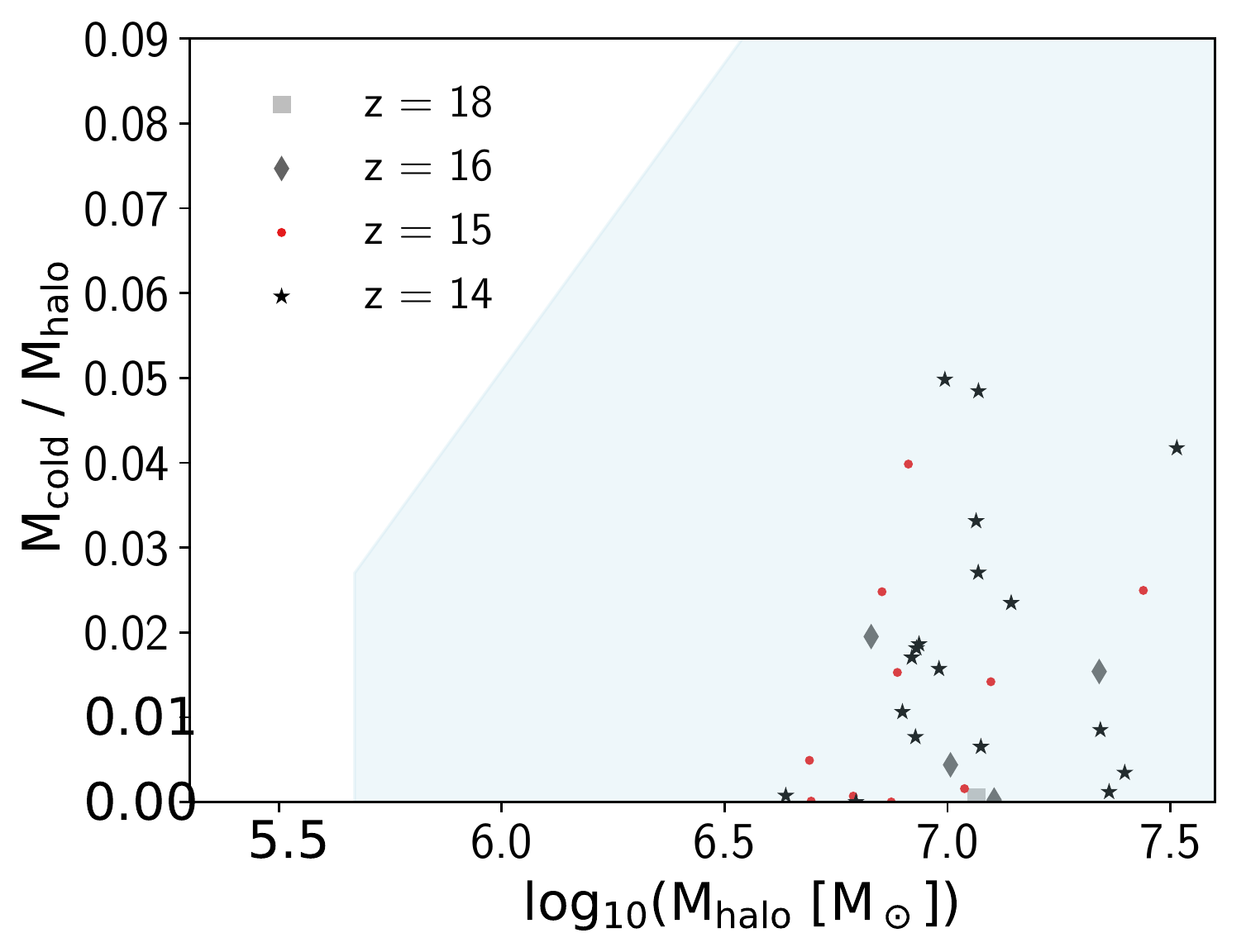}
\caption[Cold gas fraction $M_\mathrm{cold}/M_\mathrm{halo}$ as a function of
halo mass for the simulation with a streaming velocity of 2$\sigma$ for
redshifts in the range $14 \le z \le 18$.]
{Cold gas fraction $M_\mathrm{cold}/M_\mathrm{halo}$ as a function of
halo mass for the simulation with a streaming velocity of 2$\sigma$ for 
redshifts in the range $14 \le z \le 18$. 
Values are only plotted for haloes that contain at least one cold gas cell.
The shaded region shows the parameter space occupied by simulation v0.}
\label{fig:coldgasfraction_v12_allz}
\end{figure}

\begin{figure}
\centering
\includegraphics[width=0.99\columnwidth]{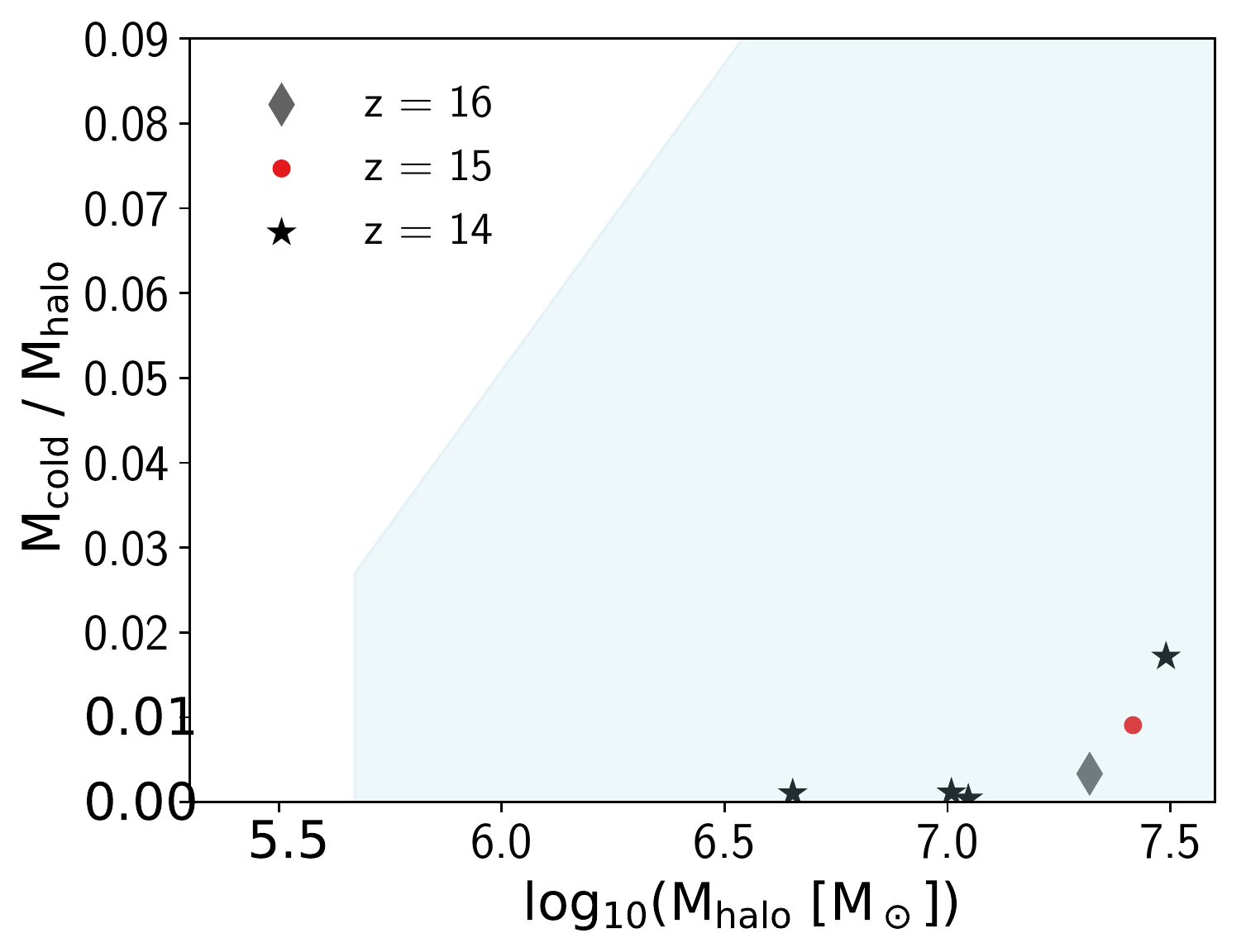}
\caption[Cold gas fraction $M_\mathrm{cold}/M_\mathrm{halo}$ as a function of
halo mass for the simulation with a streaming velocity of 3$\sigma$ in the small box for
redshifts in the range $ 14 \le z\le 16$.]
{Cold gas fraction $M_\mathrm{cold}/M_\mathrm{halo}$ as a function of
halo mass for the simulation with a streaming velocity of 3$\sigma$ in the small box for 
redshifts in the range $ 14 \le z\le 16$. 
Values are only plotted for haloes that contain at least one cold gas cell. 
The shaded region shows the parameter space occupied by simulation v0.
(Note: as there are only six data points in this figure, we increase the 
size of the markers compared to Figures \ref{fig:coldgasfraction_allz}, \ref{fig:coldgasfraction_v06_allz} and \ref{fig:coldgasfraction_v12_allz}).}
\label{fig:coldgasfraction_v18_allz}
\end{figure}

Our study goes beyond these existing studies as we are able to follow the collapse of 
gas to cold, dense structures in the centre of a halo. 
In Figures \ref{fig:coldgasfraction_v06_allz}, \ref{fig:coldgasfraction_v12_allz} 
and \ref{fig:coldgasfraction_v18_allz} 
we show the cold mass fraction $M_\mathrm{cold}/M_\mathrm{halo}$ against 
halo mass $M_\mathrm{halo}$ for the three different streaming velocity simulations. 
Compared to the case of zero streaming in Figure \ref{fig:coldgasfraction_allz}, one can immediately 
see that the number of haloes in the figures decreases with increasing streaming velocity. 
This effect cannot be reduced to the decrease in halo mass function alone. 
In regions with high streaming velocity, cold gas can only assemble in 
more massive objects than in regions with zero streaming. This can 
be seen from the same figures, as the minimum halo mass for cold 
gas shifts to higher values. 

\begin{figure*}
\centering
\includegraphics[width=1.99\columnwidth]{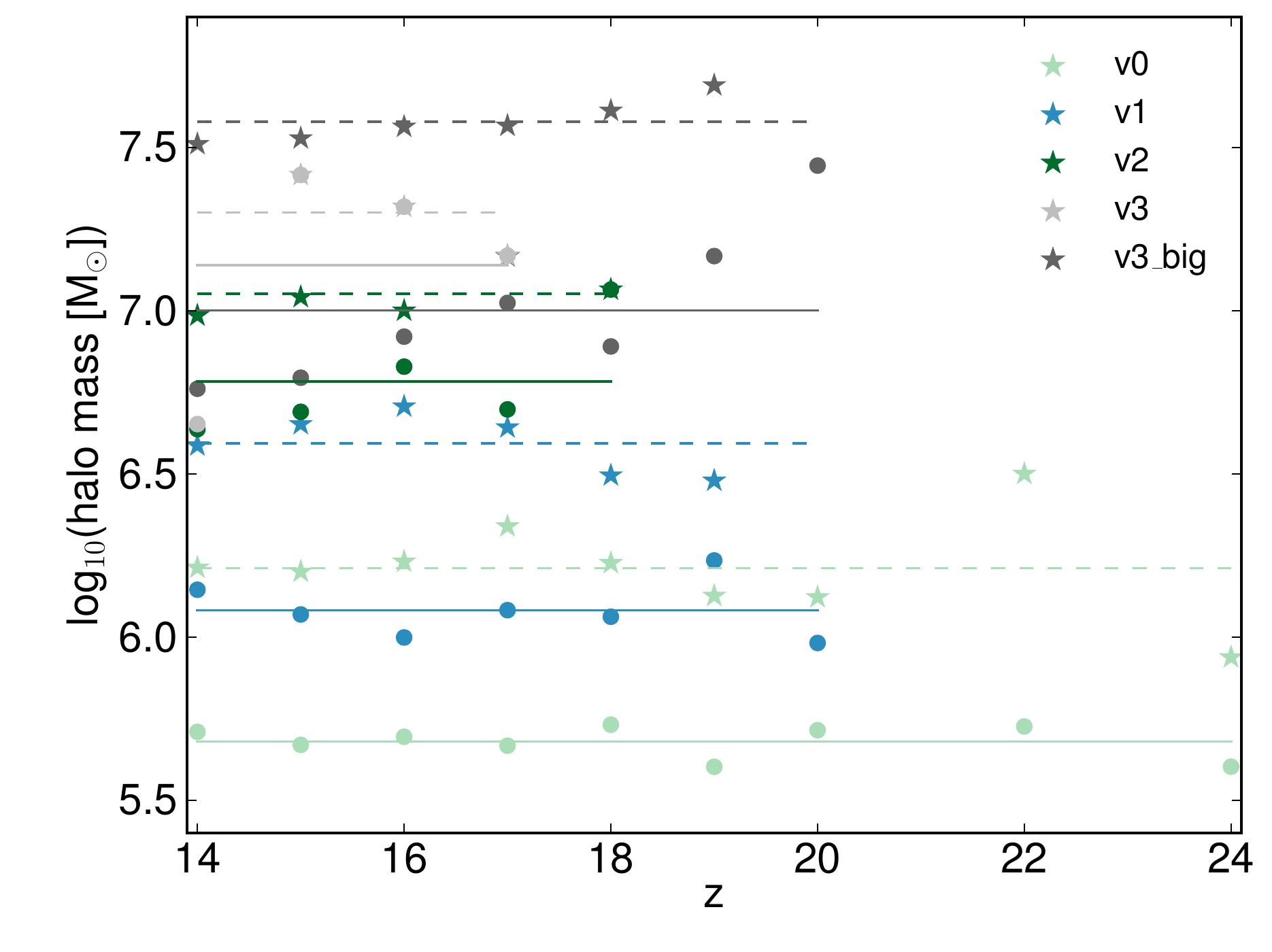}
\caption[Halo mass of the first halo to contain cold gas and halo threshold 
mass above which more than 50\% of all haloes contain cold gas as a function
of redshift for all different streaming velocities.]
{Halo mass of the first halo to contain cold gas (circles) and halo threshold 
mass above which more than 50\% of all haloes contain cold gas (stars) as a function 
of redshift. The results for all simulations are shown here. Different 
colours indicate different streaming velocities. Solid lines show the 
average mass of the first halo to contain cold gas ($M_{\rm halo, min}$) for the different streaming velocities. Dashed lines show the average halo mass above which 50\% of the haloes contain cold gas ($M_{\rm halo, 50\%}$) for the same runs.}
\label{fig:start_vs}
\end{figure*}

As in the zero streaming case, we can quantify this by determining $M_{\rm halo, min}$ and $M_{\rm halo, 50\%}$ for each simulation at each output redshift (Figure~\ref{fig:start_vs}).
We see that as the streaming velocity increases, the appearance of cold gas in the simulation is delayed: in run v0, the first cold, dense gas appears at $z \sim 26$, whereas the corresponding redshifts for runs v1, v2, and v3 are $z = 20, 18$, and 17, respectively.

Increasing the streaming velocity also increases both $M_{\rm halo, min}$ and $M_{\rm halo, 50\%}$. The increase in these values is not directly proportional to the increase in the streaming velocity, but in general, each time we increase $v_{\rm stream}$ by $\sigma_{\rm rms}$, both masses increase by a factor of a few. In Table~\ref{tab:allmasses}, we list the values of these masses, averaged over redshifts, for each simulation.

Simulation v3 
contains only one halo with cold, dense gas at redshift $z=17$\, and three more 
at our final redshift $z=14$. The apparent increase of the minimum halo mass between 
redshifts $z=17$\, and 14 is a result of this halo growing in mass and therefore of the small 
sample size. 
We therefore include the results from the large box v3\_big in Figure \ref{fig:start_vs}. Thanks to the larger simulation volume, we form many more and also more massive objects in this simulation. The first halo in this simulation to contain cold gas forms at redshift $z=20$. 

\begin{table}
\centering
\begin{tabular}{l|r|r}
Name & $M_\mathrm{halo,min}$ & $M_\mathrm{halo,50\%}$ \\
\hline
v0 & 0.48 & 1.63 \\
v1 & 1.21 & 3.93 \\
v2 & 6.08 & 11.27 \\
v3 & 13.78 & 20.02 \\
v3\_big & 10.0 & 37.9 \\
\end{tabular}
\caption{Minimum halo mass and halo mass above which more than 
50\% of all haloes contain cold gas in units of $10^6$\,\Ms.}
\label{tab:allmasses}
\end{table}

For a 1\,$\sigma_\mathrm{rms}$ streaming velocity, \citet{greif11} found an increase in the minimum mass for cooling and collapse of around a factor of three. This is in good agreement with our results: we find an increase in the minimum halo mass by a factor of $2.5$ and in the average 
halo mass by a factor of $2.4$. They also found that the onset of star formation in the halo that they studied was delayed by around $\Delta z \sim 4$. As we do not track the evolution of individual haloes in our simulations, we cannot easily compare this value with the length of time that cooling and star formation is delayed in any particular halo in our simulation. However, it is interesting to note that in run v0, cold gas is first present at $z=26$, while in run v1, cold gas is first present at $z=20$, i.e.\ the appearance of cold gas anywhere in the simulation volume is delayed by $\Delta z \sim 6$. Overall, therefore, there seems to be reasonable agreement between the results of \citet{greif11} and our own results. 
However, \citet{greif11} did not investigate the impact of higher-$\sigma$ streaming velocities. In addition, studies like \cite{greif11} and \cite{stacy11a} examined only the most massive object forming in their simulation volume, rather than the larger statistical sample examined here.

\begin{figure*}
\centering
\includegraphics[width=1.98\columnwidth]{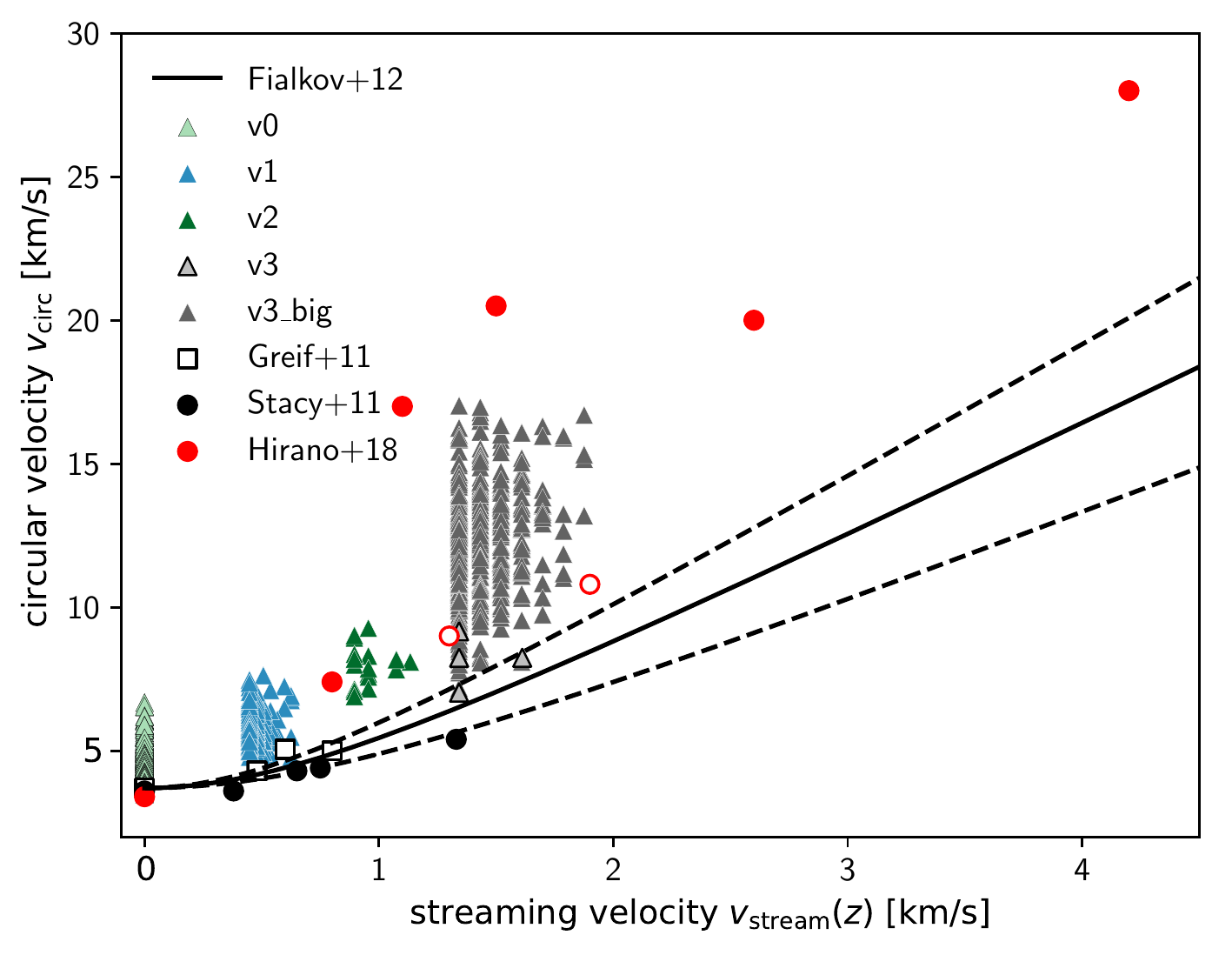}
\caption{Circular velocity $v_\mathrm{circ}{}$ of all haloes in the simulations 
as a function of the streaming velocity $v_\mathrm{stream}{}$ at the redshift of collapse. 
Data from the literature are included for comparison. 
The red circles show data from \protect\cite{hirano18}, where the open red circles 
show two collapsing minihaloes before the collapse was halted and delayed 
due to a minihalo merger and the filled circles the final collapsed haloes. 
The black, open squares show data from 
\protect\cite{greif11} and the black, filled circles show data from \protect\cite{stacy11a}.
The black lines show the corresponding fits from \protect\cite{fbth12}, where 
the solid line includes both data sets and the dashed lines  one dataset each. }
\label{fig:vcirc}
\end{figure*}

In Figure \ref{fig:vcirc}, we compare our minihaloes to literature data and the model suggested by \cite{fbth12}. We plot the circular velocity $v_{\rm circ}$ of each of our simulated haloes at the time that it first contains cold gas versus the value of the streaming velocity at the same time. We define the circular velocity for our haloes as
\begin{equation}
v_\mathrm{circ} = \sqrt{\frac{G M_\mathrm{vir}}{R_\mathrm{vir}}},
\end{equation}
where $G$ is the gravitational constant $G$, $M_{\rm vir}$ is the virial mass and $R_{\rm vir}$ is the virial radius found by the friends-of-friends algorithm. 
We evaluate the circular velocity at the redshift $z$, when a minihalo fulfills our criterion for cold, dense gas for the first time, mimicking collapse of that halo. The streaming velocity at that redshift is given by 
\begin{equation}
v_\mathrm{stream}(z) = v_\mathrm{stream,0} \left(\frac{1+z}{1+z_{\rm init}}\right),
\end{equation}
where our streaming velocities have values of 0 -- 18 km\,s$^{-1}$ at the initial redshift of $z_\mathrm{init} = 200$.

\cite{fbth12} suggest using an ``effective'' circular velocity 
threshold for star formation, based on the minimum circular velocity required for efficient cooling in the absence of streaming, $v_\mathrm{circ,0}{}$, and the streaming velocity $v_\mathrm{stream}(z)$:
\begin{equation}
v_\mathrm{circ}^{2} = v_\mathrm{circ,0}^2 + [\alpha\times v_\mathrm{stream}(z)]^2, 
\end{equation}
where $\alpha$ is a free parameter to be fit by comparison with simulation results. \cite{fbth12} find an optimal fit to the data from \cite{stacy11a} and \cite{greif11} is given by $v_\mathrm{circ,0} = 3.714$\,km\,s$^{-1}$ and $\alpha = 4.015$. We include this line 
in Figure \ref{fig:vcirc}. All of our simulations have circular velocities slightly 
higher than this ``effective'' velocity, but it gives a good lower limit 
for simulations v0, v1, v3 and v3\_big. 
Our data are in the middle of the early simulations by \cite{stacy11a} and \cite{greif11} and recent results by \cite{hirano18}. As the \cite{stacy11a} and \cite{greif11} studies select haloes that are the first objects in the simulations that collapse, it is unsurprising that we find that they lie at the bottom end of our distribution of $v_{\rm circ}$: objects in their simulations that would have required a larger $v_{\rm circ}$ to collapse would also have tended to form slightly later and hence would not have been selected. 
In the case of the \cite{hirano18} study, the halo they study is undergoing a merger process. This delays the centre from becoming dense and cold until the halo reaches higher masses. 
However, our simulation results do provide strong support for their contention that in runs with large streaming velocities, $v_{\rm circ}$ is a poor predictor of whether or not cooling and collapse can occur in a given minihalo.
\section{Conclusions}
\label{sec:stream_con}
We have performed five cosmological hydrodynamical simulations using the code 
\textsc{arepo}, including a primordial chemistry network. 
They target different regions of the Universe with zero, 1, 2 and 3 $\sigma_\mathrm{rms}$ 
streaming velocity to understand the influence of the streaming velocity value 
on first star formation. Our employed resolution is very high throughout 
the box, leading to thousands of well-resolved minihaloes. 

In a region of the Universe with zero streaming velocity, we find that in order for gas to cool within a minihalo, its mass must exceed a minimum mass scale $M_\mathrm{halo,min} \approx 4.8\times10^5$\,\Ms. However, not all minihaloes above this minimum mass contain cool gas. 
The fraction of minihaloes containing cool gas increases smoothly with increasing minihalo mass from close to zero at $M \sim M_{\rm halo,min}$ to $\sim 1$ at $M \sim 10^{7}$\,\Ms. A majority of minihaloes host cold gas for minihalo masses above $M_\mathrm{halo,50\%} \approx 1.6\times10^6$\,\Ms. 
While the minimum halo mass for cold, dense gas 
depends to some extent on our choice of density threshold, the value of $M_\mathrm{halo,50\%}$ is much more robust.
In contrast to predictions from analytical studies, we do not see a dependence 
of these characteristic mass scales on redshift. In this, our results agree well with the simulation results of \cite{yahs03}, although we find a slightly different minimum mass scale, likely due to the differences in the chemical and cooling model employed in the different simulations.

In regions of the Universe with non-zero streaming velocity, we find a substantial increase in both the minimum halo mass and the average halo mass for the formation of cold, dense gas in a halo. All these masses can be found in Table \ref{tab:allmasses} and are redshift-independent over the range of redshifts studied here. For each $ \sigma_\mathrm{rms}$ increase in the streaming velocity, we find a factor of a few increase in both $M \sim M_{\rm halo,min}$ and 
$M_\mathrm{halo,50\%}$. Importantly, this means that in regions of the Universe with streaming velocities $v_{\rm stream} \ge 3 \sigma_\mathrm{rms}$, star formation occurs almost entirely in haloes with $T_{\rm vir} > 10^{4}$\,K, the so-called ``atomic cooling haloes'' \citep[see also][]{anna17b}. 
The ``delay'' 
of star formation in regions of non-zero streaming velocities is 
purely indirect in our models due to hierarchical structure formation, 
as the more massive haloes form later. 

\cite{fbth12} have proposed that there is a simple relationship between the minimum circular velocity of a halo hosting cold gas and the streaming velocity at the time that the halo formed. 
Their model uses an ``effective'' circular 
velocity (a weighted mean of the circular velocity and the streaming velocity at the redshift of halo collapse) as a criterion for Pop~III star formation. 
Our results are broadly consistent with their prediction, if it is interpreted as a necessary but not sufficient condition for cooling: no haloes with circular velocities below their prediction host cold gas, but many haloes with circular velocities above their prediction are also devoid of cool gas. At any particular $v_{\rm stream}$, there is at least a factor of two scatter in the circular velocity required for a halo to host cool gas. At low $v_{\rm stream}$, our results are in good agreement with the simulations of \cite{stacy11a}, \citet{greif11} and \cite{hirano18}. 
But at higher $v_{\rm stream}$, we find cold gas at somewhat smaller values of $v_{\rm circ}$ than \cite{hirano18}. 
We would like to point out that \cite{hirano18} study the first halo forming in a 10\,Mpc box. This object is undergoing a merger process that halts the collapse, and the halo first grows in mass before its centre becomes cold and dense. This may explain the difference between \cite{hirano18} and our work.

Our results have important implications for the evolution of the cosmological star formation rate at very high redshifts. In the regime where most Pop~III stars are forming in minihaloes, increasing the mass threshold for efficient H$_{2}$ cooling results in a significant reduction in the star formation rate. Therefore, in this regime we would expect the star formation rate at very early times to be inversely correlated with the 
strength of the streaming velocity. A thorough exploration of the consequences of this is beyond the scope of our current paper, and so we restrict ourselves here to a single example. 

The statistical properties of minihaloes are also of interest. In the case of no streaming velocity, 
the spin distribution of present day galaxies \citep[see e.g.][]{teklu15} follows the same shape 
as the spin distribution in minihaloes \citep{souza13,mei14,hir14}. 
A rotating disk of cold, dense gas forms in the centre that is however unconnected to the angular momentum on larger scales \citep{maik18}. 
The spin of the gas increases with streaming 
velocities \citep{chiou18}, and it needs to be investigated how strongly this affects the cold star forming gas in the centre of the halo. 

The first measurement of a 21\,cm signal at high redshift \citep{bow18} 
by the EDGES experiment reveals a very strong absorption signal, hinting at coupling of gas and dark matter to cool the gas to very low temperatures at redshifts $z \sim 15-20{}$ \citep{bar18,fialkov18}. As the proposed cross-section of baryon-dark matter coupling critically depends on the relative velocity between the gas and the dark matter, streaming velocities play an important role in determining the overall effectiveness of this coupling. Regions with higher streaming velocities will have weaker coupling between baryons and dark matter, and hence slightly higher gas temperatures \citep{bar18}. However, our results demonstrate that cooling of gas in low mass minihaloes will also be strongly suppressed in these regions, resulting in a later onset of star formation and hence plausibly also a later coupling of the 21\,cm spin temperature with the gas temperature. The impact of this effect on the sky-averaged spin temperature at high redshift or on the spatial variation in $T_{\rm spin}$ in this scenario has yet to be investigated, but will be important for the interpretation not only of the current EDGES result \citep{bow18}, but also the results we expect to obtain with the next generation of telescopes, such as SKA-low or HERA. 
Detailed knowledge of the influence of streaming velocities on the formation of the first stars is a prerequisite for the correct astrophysical interpretation of the 21cm absorption signal and its implications for our understanding of the nature and properties of dark matter. 

\section*{Acknowledgments}
The authors acknowledge useful comments from the anonymous referee that helped to improve the quality of the paper.
The authors would like to thank Volker Bromm, Mattis Magg, Thomas Greif and Anastasia Fialkov for stimulating discussions. 
We are grateful to Volker Springel and his group for giving us access to and their help with the code \textsc{Arepo}. 
Support for this work was provided by NASA through the NASA Hubble Fellowship grant 
HST-HF2-51418.001-A   
awarded  by  the  Space  Telescope  Science  Institute,  which  is  operated  by  the  
Association  of  Universities for  Research  in  Astronomy,  Inc.,  for  NASA,  under  contract  
NAS5-26555.
All authors acknowledge support from the European Research Council under the European Community's Seventh Framework 
Programme (FP7/2007 - 2013) via the ERC Advanced Grant ``STARLIGHT: Formation of the 
First Stars" (project number 339177). 
SCOG and RSK also acknowledge support from 
the Deutsche Forschungsgemeinschaft via SFB 881, ``The Milky Way System'' (sub-projects
B1, B2 and B8) and SPP 1573 , ``Physics of the Interstellar Medium'' (grant number GL 668/2-1).
The authors gratefully acknowledge the Gauss
Centre for Supercomputing for funding this project by providing
computing time on the GCS Supercomputer SuperMUC at the Leibniz Supercomputing Centre under projects pr92za and pr74nu. 
The authors acknowledge support by the state of Baden-W\"urttemberg through bwHPC and the German Research Foundation (DFG) through grant INST 35/1134-1 FUGG. 
\bibliographystyle{mn2e}
\setlength{\bibhang}{2.0em}
\setlength\labelwidth{0.0em}
\bibliography{refs}

\appendix

%
%

\section{Updates to the chemical model}
\label{app:chem}
The primordial chemistry model used in this study is based on the model implemented in
{\sc arepo} by \citet{hartwig15a}, but has been updated in several respects. The most significant
difference is the inclusion of a simplified treatment of deuterium chemistry designed to track the
formation and destruction of HD. Our treatment of the deuterium chemistry follows that in 
\citet{cgkb11} and full details can be found in that paper. In the interests of brevity, we do not repeat
them here. 

The second update involves our treatment of the dielectronic recombination of ionized helium, He$^{+}$.
In \citet{hartwig15a} we used a rate for this process taken from \citet{ap73}. In our present study, we
use instead the more recent determination by \citet{b06}. In practice, we expect this change to have no 
significant  impact on our results, as at the gas temperatures probed in this study, very little ionized helium
is present, and the small amount that is created recombines primarily via radiative recombination rather than
dielectronic recombination. 

Finally, we have also updated our treatment of H$^{-}$ photodetachment. In all of our runs, we account for 
photodetachment of H$^{-}$ by CMB photons. The rate for this process is given by the sum of two contributions,
one describing the influence of the thermal CMB photons and a second describing the influence of the non-thermal
radiation background produced by cosmic recombination at $z \sim 1100$ \citep{hp06}. For the thermal contribution, 
we adopt the following rate coefficient from \citet{gp98}, 
\begin{equation}
R_{\rm H^{-}, th} = 0.11 T_{\rm r}^{2.13}  \exp \left(\frac{-8823\mathrm{K}}{T_{\rm r}} \right) \: {\rm s^{-1}} ,
\end{equation}
where $T_{\rm r} = T_{\rm CMB} = 2.73 (1 + z)$~K. For the non-thermal contribution, we adopt the expression
\citep{cop11}
\begin{equation}
R_{\rm H^{-}, nth} = 8 \times 10^{-8} T_{\rm r}^{1.3} \exp \left(\frac{-2300\mathrm{K}}{T_{\rm r}} \right) \: {\rm s^{-1}}.
\end{equation}
The total H$^{-}$ photodetachment rate is simply the sum of these two terms.

\section{Numerical convergence}
\label{converge}
We want to provide robust results that are numerically converged. Therefore, we test 
our fiducial resolution of $2\times1024^3$ particles in a (1\,cMpc/$h$)$^3$ 
volume against larger and smaller mass resolutions 
with otherwise identical initial conditions. For simulations with box length of 
500\,ckpc/$h$, we vary the number of particles 
from $2 \times 128^3{}$ to $2\times 1024^3{}$ in steps of eight, so that the mass resolution changes between each simulation by a factor of 8. 
For our convergence study, we consider the case with zero streaming velocities, as this run has the lowest halo mass threshold for efficient H$_{2}$ cooling. If the lowest mass haloes containing cold gas are well resolved in this simulation, the more massive haloes containing cold gas in the simulations with non-zero streaming velocities should certainly be adequately resolved.

\begin{figure}
\centering
\includegraphics[width=0.99\columnwidth]{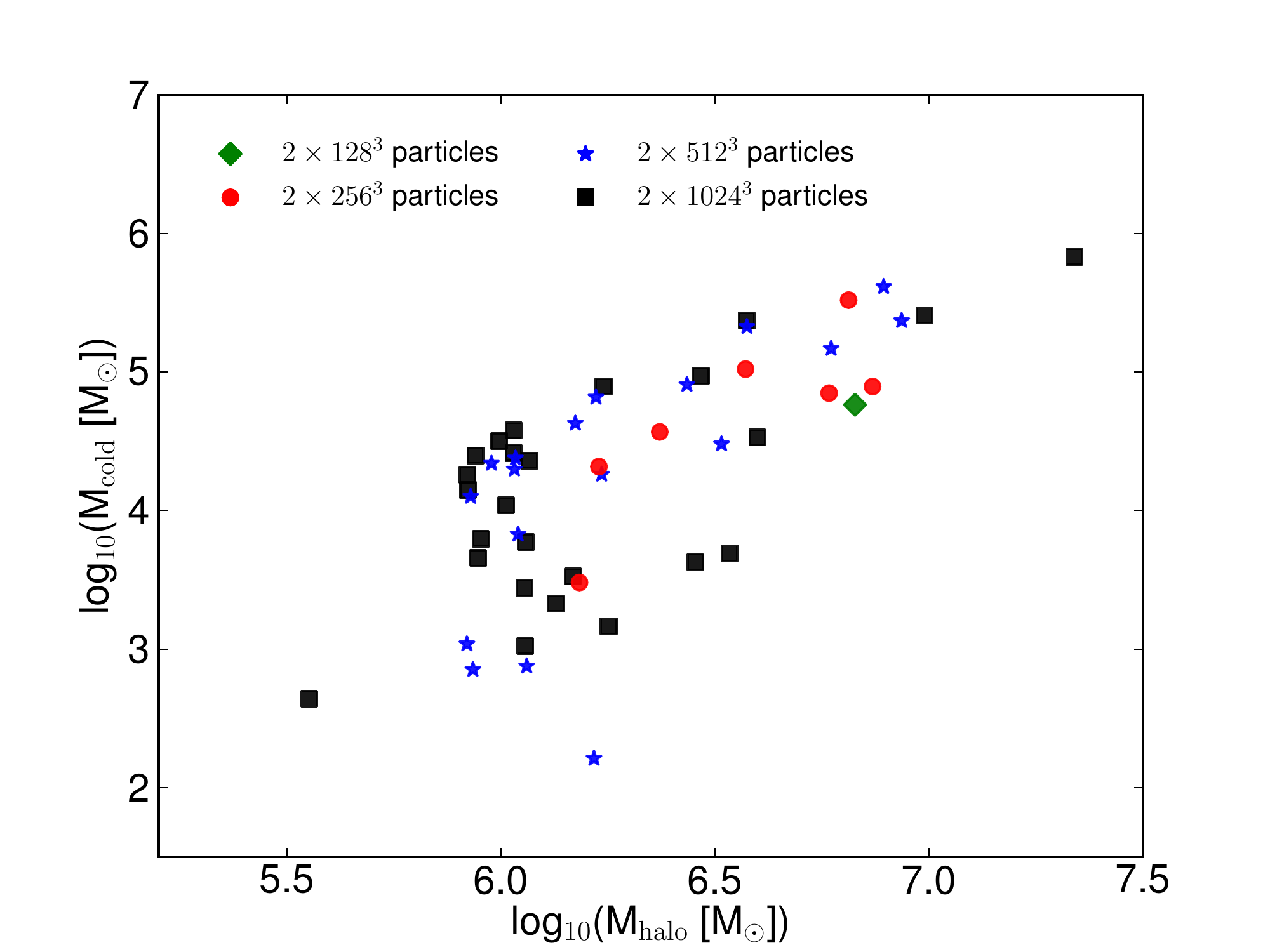}
\caption[Resolution test: different number of particles and Voronoi cells for a 
cosmological box with the side length of 500\,ckpc$/h$.]
{Resolution test: different number of particles and Voronoi cells for a 
cosmological box with the side length of 500\,ckpc$/h$. There are many fewer
red circles ($2\times256^3$ particles) and only one green diamond ($2\times128^3$ 
particles) than blue stars ($2\times512^3{}$ particles) or black squares
($2\times1024^3{}$ particles). Except for one halo at $\approx 10^{5.6}$\,\Ms, the onset 
of efficient cooling at a halo mass $\approx 10^{5.8}$\,\Ms is the same for the 
$2\times512^3{}$ and $2\times1024^3{}$ runs.}
\label{fig:res}
\end{figure}
In Figure \ref{fig:res} we show the cold gas mass against halo mass for the four
realizations. One can see that for the two lower resolutions of $2 \times 128^3$ 
particles and $2 \times 256^3$ particles, the cold gas mass is lower than 
for the two higher resolution cases. Our fiducial resolution of $2 \times 512^3$
particles (which is equivalent to $2 \times 1024^3$ particles in our volume of 1\,cMpc$^3{}$ in the production runs) agrees well with a even higher resolution of $2 \times 1024^3$ particles. 
We are therefore confident that our simulations are well converged. 

These results also help to motivate our choice of simulation volume in our large box simulation, v3\_big. They demonstrate that in order to properly resolve cooling in these haloes, we to resolve each halo with at least a few thousand dark matter particles. In our large box simulation, low mass haloes with $M_{\rm halo} \sim 10^{6} \: {\rm M_{\odot}}$ are resolved with only around a hundred particles, and so cooling is not modelled accurately in these haloes. However, we know from run v3 that the minimum mass required for cooling in a simulation with a $3\sigma$ streaming velocity is increased by over an order of magnitude, to $M \sim 10^{7} \: {\rm M_{\odot}}$, and haloes of this mass are resolved by at least a thousand particles. Therefore, the lowest mass haloes in which efficient H$_{2}$ cooling occurs in run v3\_big are resolved by a comparable number of dark matter particles to the lowest mass haloes in which efficient H$_{2}$ cooling occurs in run v0.
%
%
\section{Comparison of friends-of-friends results with subfind}
\label{app:fof}
In the high-redshift Universe, minihaloes tend not to be spherical 
but instead are very elongated and irregular \citep[see e.g.][]{mei14}.
Our results could therefore depend on the halo finding
algorithm. In this Appendix, we provide a check to show that 
our results are independent of the halo selection algorithm. 
In this paper, we have considered all cold gas particles that were associated to a halo by the 
standard friends-of-friends (fof) algorithm. 
Another possibility is to consider only gas that belongs to the most bound subhalo 
in the fof structure. For this test, we use the halo finding algorithm subfind \citep{subfind}.

\begin{figure}
\centering
\includegraphics[width=0.99\columnwidth]{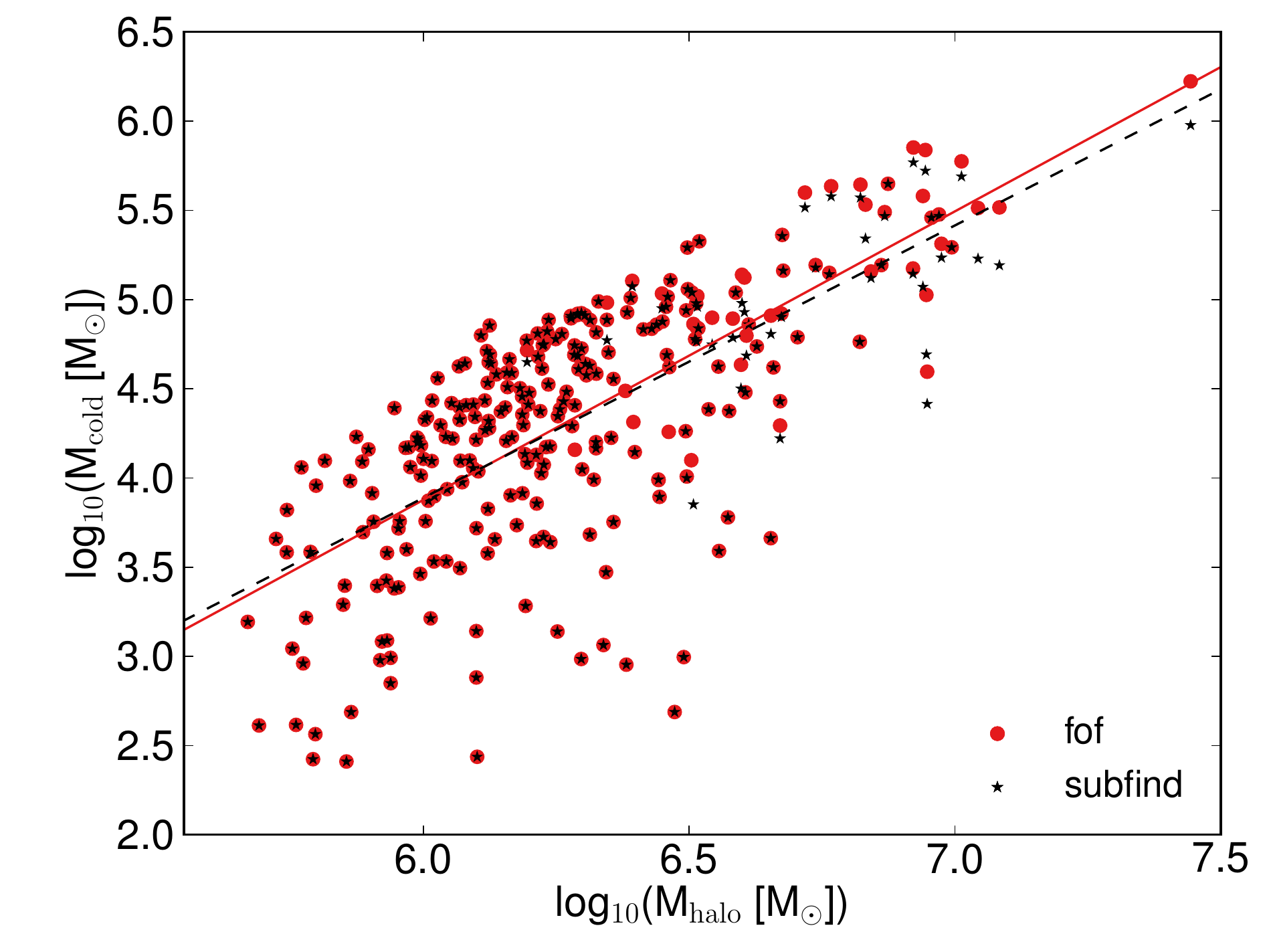}
\caption[Cold gas mass against halo mass for the simulation without streaming
at redshift $z=15$: comparison of fof with subfind.]
{Cold gas mass against halo mass for the simulation without streaming 
at redshift $z=15$. 
We show the cold gas mass selected by two different halo finders: 
cold gas found in the whole friends-of-friends (fof) structure (red circles) and cold gas only found in the most massive subhalo (black stars). In most cases, there is very good agreement, but in the most massive haloes, the fof halo finder tends to recover slightly larger cold gas masses than subfind.}
\label{fig:subf}
\end{figure}
In Figure \ref{fig:subf}, we show the cold gas mass as a function of halo mass for the 
simulation without streaming at redshift $z=15$ (red circles). 
Overplotted are the cold gas masses in the most bound subhaloes (black stars). 
The cold gas mass found in the most bound 
subhalo agrees very well with the cold gas mass found in the whole 
fof structure. At the highest halo masses, some haloes contain 
more cold gas when considering the whole fof structure instead of only the most bound subhalo, but this difference does not affect any of the main results of the paper. We therefore conclude that for this study, our results are not sensitive to our choice of halo finding algorithm. 
%
%
\section{Cold mass -- halo mass relation}
\label{sec:lw2_relation}
In Section \ref{sec:stream_ana1}, 
we have provided a fit for the cold gas mass -- halo mass relation. 
As can be seen in Figure \ref{fig:mhalo_mcold_z15}, the scatter is significant  
and should not be neglected when using our fitting approximation. 
Despite this caveat, we provide fits to all our simulations in Table \ref{tab:allfits}
that could be used for semi-analytical calculations. This relation is valid only for the mass range of haloes formed in these simulations, 
from about $5 \times 10^5$\,\Ms\, to $3 \times 10^7$\,\Ms\, in most cases. 
The data in this table is given in the form 
\begin{equation}
\log_{10} \left( M_\mathrm{cold} \right) = a \times \log_{10} \left( M_\mathrm{halo}\right) - b .
\end{equation}
\begin{table}
\begin{center}
\begin{tabular}{l|cccc}
Name  & $a$ & $b$ & $a$ & $b$\\
$z$      & 15& 15  & 20 & 20\\
\hline
v0      & 1.6 &  5.8 & 1.3 &  4.0 \\   
v1      & 1.8 &  7.2 & 1.2 &  3.8 \\
v2      & 3.1 & 17.0 & --  &  --  \\
v3\_big & 1.7 &  7.7 & 2.8 & 16.0 \\
\hline
\end{tabular}
\caption[Cold gas mass -- halo mass relation for all simulations at redshifts 15 and 20.]
{Cold gas mass -- halo mass relation for all simulations at redshifts 15 and 20. Simulation v2 does not have enough data points at $z=20$ to provide a meaningful fit. The same is true for simulation v3 at both redshifts, and so we omit it from the table.}
\label{tab:allfits}
\end{center}
\end{table}
%
\label{lastpage}

\end{document}